\documentclass[sigconf,screen]{acmart}
\renewcommand\footnotetextcopyrightpermission[1]{}
\setcopyright{none}
\acmConference[Information and Software Technology]{}{IST Journal 2021}{Elsevier}

\usepackage{amsmath}
\usepackage{graphicx}
\usepackage{float}
\usepackage{pbox}
\usepackage{url} 
\usepackage{hyperref}
\usepackage{fancybox}
\usepackage{multirow}
\usepackage{xspace}
\usepackage{adjustbox}
\usepackage{xcolor}
\usepackage{comment}

\usepackage[shortlabels]{enumitem}
\usepackage[labelfont=bf,textfont=md]{caption}

\usepackage{listings}
\usepackage{soul}
\newcommand{\floor}[1]{\lfloor #1 \rfloor}

\usepackage{array}
\newcolumntype{L}[1]{>{\raggedright\let\newline\\\arraybackslash\hspace{0pt}}m{#1}}
\newcolumntype{C}[1]{>{\centering\let\newline\\\arraybackslash\hspace{0pt}}m{#1}}
\newcolumntype{R}[1]{>{\raggedleft\let\newline\\\arraybackslash\hspace{0pt}}m{#1}}

\usepackage{balance}
\usepackage[ textsize=tiny, textwidth=1.4cm]{todonotes}

\newcommand{\npa}{neural program model\xspace}
\newcommand{\npas}{neural program models\xspace}

\newcommand{\Npas}{Neural program models\xspace}

\newcommand{\NPAS}{Neural Program Models\xspace}

\newcommand{\Fix}[1]{\textbf{\textcolor{red}{Fix/TODO}: #1}}

\newcommand{\Space}[1]{}

\usepackage{xspace}

\newcommand{\Comment}[1]{}
\newcounter{observation}
\newcommand{\observation}[1]{\refstepcounter{observation}
        \begin{center}
        \Ovalbox{
        \begin{minipage}{0.93\columnwidth}
                \textbf{Observation \arabic{observation}:} #1
        \end{minipage}
        }
        \end{center}
}

\newcommand{\Part}[1]{\noindent\textbf{#1}}

\newcommand{\ctv}{\emph{code2vec}\xspace}
\newcommand{\cts}{\emph{code2seq}\xspace}
\newcommand{\ggnn}{\emph{GGNN}\xspace}

\newcommand{\VN}{\emph{Variable Renaming}\xspace}
\newcommand{\BX}{\emph{Boolean Exchange}\xspace}
\newcommand{\LX}{\emph{Loop Exchange}\xspace}
\newcommand{\PS}{\emph{Permute Statement}\xspace}
\newcommand{\SF}{\emph{Switch to If}\xspace}

\newcommand{\UN}{\emph{Unused Statement}\xspace}

\newcommand{\JS}{\textsc{Java-Small}\xspace}
\newcommand{\JM}{\textsc{Java-Med}\xspace}
\newcommand{\JL}{\textsc{Java-Large}\xspace}

\newcommand{\eg}{\textit{e.g.}\xspace}
\newcommand{\ie}{\textit{i.e.}\xspace}
\newcommand{\etal}{\textit{et al.}\xspace}

\title[On the Generalizability of \NPAS w.r.t. Semantic-Preserving Program Transformations]{On the Generalizability of \NPAS with respect to Semantic-Preserving Program Transformations}

\author{Md Rafiqul Islam Rabin}
\affiliation{University of Houston}

\author{Nghi D. Q. Bui}
\affiliation{Singapore Management University}

\author{Ke Wang}
\affiliation{Visa Research}

\author{Yijun Yu}
\affiliation{The Open University}

\author{Lingxiao Jiang}
\affiliation{Singapore Management University}

\author{Mohammad Amin Alipour}
\affiliation{University of Houston}

\keywords{neural models, code representation, model evaluation, program transformation, generalizability}

\begin{document}
\begin{abstract}
\textbf{Context}: With the prevalence of publicly available source code repositories to train deep neural network models, \npas can do well in source code analysis tasks such as predicting method names in given programs that cannot be easily done by traditional program analysis techniques.
Although such \npas have been tested on various existing datasets, the extent to which they generalize to unforeseen source code is largely unknown.
\textbf{Objective}: Since it is very challenging to test \npas on all unforeseen programs, in this paper, we propose to evaluate the generalizability of \npas with respect to semantic-preserving transformations: a generalizable \npa should perform equally well on programs that are of the same semantics but of different lexical appearances and syntactical structures.
\textbf{Method}: We compare the results of various \npas for the method name prediction task on programs before and after automated semantic-preserving transformations.
We use three Java datasets of different sizes and three state-of-the-art neural network models for code, namely \emph{code2vec}, \emph{code2seq}, and \emph{GGNN}, to build nine such \npas for evaluation. 
\textbf{Results}: Our results show that even with small semantically preserving changes to the programs, these \npas often fail to generalize their performance.
Our results also suggest that \npas based on data and control dependencies in programs generalize better than \npas based only on abstract syntax trees (ASTs). 
On the positive side, we observe that as the size of the training dataset grows and diversifies the generalizability of correct predictions produced by the \npas can be improved too.  
\textbf{Conclusion}: Our results on the generalizability of \npas provide insights to measure their limitations and provide a stepping stone for their improvement.

\end{abstract}

\begin{comment}
In recent years, with the prevalence of publicly available source code repositories and  advances in machine learning algorithms, especially deep neural networks, data-driven approaches have made significant progresses in source code analysis.
These approaches, called neural program analyzers, use neural networks to train models to make  predictions on given programs for tasks ranging from development productivity to program reasoning.  
Despite the growing popularity of neural program analyzers, the extent to which their results generalizable to new, unseen source code is unknown. 
In many cases\Fix{some number here} the neural program analyzers fail to generalize well, sometimes to programs with negligible textual differences.
\end{comment}

\maketitle
\renewcommand{\shortauthors}{M.R.I. Rabin, N.D.Q. Bui, K. Wang, Y. Yu, L. Jiang, M.A. Alipour}

\section{Introduction}

Abundance of publicly available source code repositories has enabled a surge in data-driven approaches to programs analysis tasks.
Those approaches aim to discover common programming patterns for various downstream applications \cite{BigCodeSurvey} that are not easily achievable via traditional program analysis techniques,
% various program analysis and software engineering applications, 
\eg, prediction of data types in dynamically typed languages~\cite{vincent:type}, detection of the variable naming issues~\cite{allamanis2017learning}, or repair of software defects~\cite{dinella2019hoppity}. 
The advent of deep neural networks has accelerated the innovation in this area and has greatly enhanced the performance of these approaches. 
The performance of deep neural networks in cognitive tasks such as method name prediction or variable naming has reached or exceeded the performance of other data-driven approaches. 
The performance of neural networks has encouraged researchers to increasingly adopt neural networks in program analysis tasks, giving rise to increasing uses of \npas.

While the performance of \npas continues to improve, the extent to which they can generalize to new, unseen programs is still unknown, even if the programs are in the same programming language. 
This problem is of more importance if we want to use them in downstream safety-critical tasks, such as malware detection and automated defect repair.
This problem is particularly hard, as the interpretation of neural models that constitute the core reasoning engine of \npas remains challenging---especially for the complex neural networks (\eg, RNN) that are commonly used in the proposed \npas.

A comprehensive understanding of the extent of generalizability of \npas would help developers to know when to use data-driven approaches and when to resort to traditional deductive methods of program analysis.
It would also help researchers to focus their efforts on devising new techniques to alleviate the shortcomings of existing \npas.
Lack of knowledge about the limits of \npas may exaggerate their capability and cause careless applications of the \npas on the domains that they are not suited for; or, spending time and efforts on developing \npas while a traditional, more understandable technique can perform equally well or better.

Recently, we have seen a growing interest in the rigorous evaluation of \npas.
\citet{Wang:2019:Coset} compared the robustness of different program representations under compiler optimization transformations. They found that the program representations based on static code features are more sensitive to such changes than dynamic code features.
\citet{Milto:Onward:2019} evaluated the impact of code duplication in various \npas and found that code duplication in the training and test datasets inflated the performance of almost all current \npas. 
More recently, preliminary studies in this field started to emerge; \eg,
\citet{rabin2019tnpa,rabin2020evaluation} proposed the idea of testing \npas using semantic-preserving transformations;
\citet{Nghi2019AutoFocus} measured the impact of a specific code fragment by deleting it from the original source code;
\citet{zhang2020generating} proposed a sampling approach to generate adversarial examples for code classification models; and \citet{compton2020embedding} showed that the obfuscation of variable names makes a model on source code more robust with less bias towards variable names.
Further, \citet{yefet2019adversarial} followed and proposed adversarial example generation for \npas using prediction attribution \cite{Attribution};
\citet{Reps:CodeRobustness} increased robustness of neural representations of code by adding semantically equivalent programs to the training data; and
\citet{Vechev:AdversarialCode} proposed an approach for increasing the robustness of \npas for type prediction based on finding prediction attribution, adversarial training, and refining source code representations.
Although these studies share the similar ultimate goal of evaluating and improving the performance of \npas with respect to unseen programs, 
there is still a lack of systematic quantifiable metrics to measure the extent to which the \npas can generalize to unseen programs,
and it would not be fair either to evaluate a \npa against all possible unseen programs that it was not designed for.

\Part{Goal.}
In this paper, we attempt to understand the limits of \emph{generalizability} of \npas by comparing their behavior before and after semantic-preserving program transformations.
That is, how the results of a \npa generalize to a semantically-equivalent program.
By limiting unseen programs to semantically equivalent ones and controlling the semantic-preserving program transformations, we are able to provide a fair, systematic, quantifiable metric for evaluating the generalizability of a \npa.

In this paper, we report the results of a study on the generalizability of three highly-cited \npas: \ctv~\cite{alon2018code2vec}, \cts~\cite{alon2018code2seq}, and \ggnn~\cite{fernandes2018structured}.
To evaluate their generalizability, we transform programs in the original datasets for testing to generate semantically-equivalent counterparts. 
We employ six semantic-preserving transformations that impact the structure of programs (\ie abstract syntax trees) with varying degrees, ranging from common refactoring, \eg, variable renaming, to more intrusive changes such as changing for-loops to while-loops.

Our results suggest that all \npas evaluated in this study are sensitive to the semantic-preserving transformations; that is,
the output of the \npa would be different on transformed programs compared to its output on the original programs.
This sensitivity remains an issue even in the cases of small changes to the programs, such as renaming variables or reordering independent statements in a basic block. 
Moreover, our results suggest that \npas (\eg, \ggnn) that encode data and control dependencies in programs generalize better than the \npas that are solely based on abstract syntax trees, and in most cases the generalizability of a \npa can be improved with the growth in the size of training datasets.

The results of this study reveal that the generalizability of \npas is still far from ideal and require more attention from the research community to devise more generalizable models of source code, or designing pre-processing techniques, \eg canonicalizing program representations, to increase immunity of \npas to such program transformations. 
% \ke{The following sentence is redundant}
% The result signifies that we need to go beyond the mere application of neural networks and devise specialized techniques and methodologies to analyze neural models that process source code.

Compared to closely related work by \citet{yefet2019adversarial} and \citet{Reps:CodeRobustness} where their goals are adversarial code generation and increasing robustness of \npas, 
this paper provides a complementary view to the evaluation of \npas by focusing on the evaluation of generalizability of \npas with a large number of transformations, and in-depth analysis of changes in their behavior on transformed programs. 
This paper also evaluates the impact of the size of datasets and programs on the generalizability of \npas. 
%\citet{Yefet:Arxiv:2019} propose generation of adversarial examples for \npas with two code transformations. In this paper, we evaluate six transformations Their work fit does not provide generalization, and \citet{Reps:CodeRobustness} explores increasing the robustness of the networks by retraining the \npas by transformed programs without providing analysis of .

%Perhaps the closest work to this work is ~\citet{Yefet:Arxiv:2019} where it develops adversarial attack using gradient-based algorithm to locate and modify parts of the input code to impact the prediction of of the model.
%However, the adversarial attacks are only limited to variable names (i.e. rename existing variables or add new unused variables), and do not consider any non-variable transformations such as permute independent statements or replace for/while loops. As models rely on variable names for prediction~\cite{compton2020embedding}, models could be easily altered by modifying variable names or adding new variables, but could become more robust against them by ignoring variable names or filtering outlier variable names for prediction. Another concurrent work by \citet{Reps:CodeRobustness} that mainly focuses on making a model more robust upon training with semantic adversarial examples, but only limited to sequence-based models of source code (i.e. seq2seq and code2seq).
\smallskip

\Part{Contributions.} This paper makes the following contributions.
\begin{itemize}[nosep,leftmargin=*]
	\item We introduce the notion of generalizability with respect to semantic-preserving transformations for \npas.
	\item We perform a large-scale study to evaluate the generalizability of state-of-the-art \npas. We also provide insights into the generalizability of existing \npas and discuss their practical implications.
	\item We provide an in-depth analysis of changes in the prediction and evaluate the impact of the size of datasets and programs on the generalizability of \npas.
\end{itemize}

\begin{comment}
\smallskip
The rest of the paper is organized as follows.
Section~\ref{sec:example} presents a motivating example and provides our definition of generalizability.
Section~\ref{sec:background} provides an introduction to \npas.
Section~\ref{sec:approach} presents our approach for evaluating the generalizability of a \npas.
Sections~\ref{sec:settings} and~\ref{sec:results} describe our evaluation settings and actual results.
Section~\ref{sec:discuss} discusses practical implications of our results.
Section~\ref{sec:related} compares with closely related work.
Section~\ref{sec:threats} discusses threats to validity of this study.
Section~\ref{sec:conclude} concludes with future work.

\end{comment}

\section{Motivating Example \& Definition}
\label{sec:example}

% \subsection{Motivating Example}

We use \ctv~\cite{alon2018code2vec} for exposition in this section.
The \ctv~\cite{alon2018code2vec} is a recent, highly-cited (200+ citations as of Nov.~2020) \npa that predicts the name of a Java method given the body of the method.
Such a \npa can assist developers in classification of methods, code similarity detection, and code search.

%\subsubsection{Variable Renaming}

\begin{figure*}[t!]
\noindent \begin{minipage}{.5\textwidth}
\begin{lstlisting}[title= Prediction before transformation: \textbf{\textcolor{blue}{compareTo}}]
public int compareTo(ApplicationAttemptId <@\textbf{\textcolor{blue}{other}}@>) {
    int compareAppIds = this.getApplicationId()
        .compareTo(<@\textbf{\textcolor{blue}{other}}@>.getApplicationId());
    if (compareAppIds == 0) {
        return this.getAttemptId() - <@\textbf{\textcolor{blue}{other}}@>.getAttemptId();
    } else {
        return compareAppIds;
    }
}
\end{lstlisting}
\end{minipage}%
\begin{minipage}{.5\textwidth}
\begin{lstlisting}[title= Prediction after transformation: \textbf{\textcolor{red}{getCount}}]
public int compareTo(ApplicationAttemptId <@\textbf{\textcolor{red}{var0}}@>) {
    int compareAppIds = this.getApplicationId()
        .compareTo(<@\textbf{\textcolor{red}{var0}}@>.getApplicationId());
    if (compareAppIds == 0) {
        return this.getAttemptId() - <@\textbf{\textcolor{red}{var0}}@>.getAttemptId();
    } else {
        return compareAppIds;
    }
}
\end{lstlisting}
\end{minipage}
% \vspace{-1em}
\caption{\VN on \texttt{java-small/test/hadoop/ApplicationAttemptId.java} file.}
\label{fig:motivating-example}
% \vspace{-1em}
\end{figure*}

\begin{comment}
\begin{figure*}[t!]
    \centering
    \includegraphics[width=\columnwidth]{motivating-example/m_example.png}
    \caption{\VN on \texttt{java-small/test/hadoop/ApplicationAttemptId.java} file.}
    \label{fig:motivating-example}
\end{figure*}
\end{comment}

Figure~\ref{fig:motivating-example} shows two semantically-identical methods that implement \texttt{compareTo} functionality. The only difference between them is in the name of one of the variables.
The left snippet in Figure~\ref{fig:motivating-example} uses \texttt{other}, while the code on the right uses \texttt{var0}\footnote{\texttt{var0} is not an uncommon identifier name in Java as it appears in the training vocabulary of the datasets. At the time of writing, a search on the GitHub returns more than 75K Java classes that use this identifier.}.
However, the \ctv outputs, \ie, predictions, on these semantically equivalent programs are drastically different. \ctv predicts the snippet on the left to be \texttt{compareTo} function, and the function on the right to be \texttt{getCount}.
It seems that the predictions of \ctv rely much on the identifier names (\eg, \texttt{other}). 
This reliance would make \ctv susceptible to a common refactoring such as variable renaming, and would make it not generalize to the code snippets that are semantically the same, but are different syntactically, even under common transformations.

Lack of generalizability would lead to distrust in the \npas and hamper their wider adoption and application.
If such \npas were to be deployed in the problem settings wherein higher levels of generalizability are required,
\eg, malware detection and bug repair,
it would be much better for the \npas to demonstrate a high level of generalizability with respect to certain metrics.

\smallskip

\Part{Generalizability.} We define generalizability as the capability of a \npa to return the same
%\Fix{Amin:same or similar, or some other word? LX: it sounds like it should be "the same", at least for semantic-equivalent programs.} 
results under semantic-preserving transformations.
%\footnote{

%\Fix{the footnote sounds important, at least as important as the next sentence "We also note that..." to  to differentiate our work from related work; could put the footnote into the main text; also, it avoids a big gap latex puts before the paragraph.}
In this paper, we differentiate {\em generalizability} from the term \emph{robustness} that is commonly used in the neural network literature~\cite{szegedy2013intriguing} for two main reasons. First, robustness is usually defined in the face of adversarial examples that have security implications, while we do not generate adversarial examples.
Second, robustness implies \emph{imperceptible} differences in the two focal inputs (\eg, minor pixel changes in two images) that are hard to attain in a sparse domain such as source code; the program transformations used in this paper often lead to perceptible changes of textual appearances and syntactic structures in program code. 
%}
We also note that our definition of generalizability differs from what is used in~\cite{Kang:ASE:2019:Generalizability}
that evaluates the usefulness of a \npa in various downstream tasks.
%, while we evaluate their generalizability to semantically-equivalent programs.
%Moreover, the generalizability in this work, while related to, is not the same as neural {\em robustness}~\cite{szegedy2013intriguing}, as robustness requires imperceptible changes to input data that may be considered as adversarial examples and has implications in reliability and security.
%Although the impact of our transformations on the semantic of the program is imperceptible, the changes to the textual and syntactic structures of the program can be perceptible.

%\Fix{Amin: we have to rewite the rest of this para.}
Together with clearly defined semantic-preserving program transformations and their change impact on the prediction results of \npas (cf.~Section~\ref{sec:approach}), we aim to provide a systematic quantifiable way to measure the generalizability of \npas, and thus shed lights on their capabilities and limits for future improvements.
With the extensibility of the program transformations and the measurements of their change impact, our evaluation approach may also be extended to measure the generalizability of \npas more comprehensively in the near future.
%Despite the significant progresses made in novel application of neural networks for program analysis tasks, their generalizability with respect to program transformations have not been adequately explored. 
%For example, a task such as malware detection would benefit greatly from a robust  from a tool that is code obfuscation would benefit from functionalities robust \npas.  generalizability of the 

\section{Background}
\label{sec:background}
Most \npas use neural network classifiers in their core components that take a code snippet or a whole program as an input, and make predictions about some of its characteristics; \eg, a bug prediction classifier that predicts the buggy-ness of statements in the input program.

Performance of a \npa depends on three main factors: quality of data (\ie, source code for this study), the representation of data for the neural network, and the neural network characteristics and its training parameters.
%\Fix{Amin: do we want to move the discussion about feature learning in the discussion to here?}

\begin{comment}
     Since neural networks can only process vectors of numbers, discrete-value data such as natural language and source code need to be represented as a vector of numbers. The features in the source code that is used to create that vector representation can impact the performance of a \npa.
    Finally, the characteristics of the neural networks---\eg, type, topology, and hyper-parameters---used in a \npa influences its performance.
\end{comment}

Quality of the data is concerned with the representativeness of data, and proper cleaning and preprocessing of the data.
Currently, most studies use open-source projects usually in mainstream programming languages, \eg, C\#, Java, C, or JavaScript. 
The available datasets for these tasks are still very immature and not standardized, and their quality is somewhat unknown.
For example, a recent study by Allamanis~\cite{Milto:Onward:2019} showed that virtually all available datasets suffer from code duplication that can greatly impact the performance of \npas.

The second factor affecting the performance of \npas is source code representations. Since neural networks need to take vectors of numbers as direct inputs, source code embeddings are used to produce a vector representation of source code.
The representation determines which program features to include and how they should be represented in the vector embeddings. The representations can be broadly categorized into two categories: static and dynamic. Static program representations consider only the features that can be extracted from parsing texts of the programs, while dynamic representations include some features pertaining to the real executions of the programs. 

The third factor impacting the performance of a \npa is the characteristics---\eg, type, topology, and hyper-parameters---of the neural networks it uses.  There are numerous choices of network architectures each with different properties.
Currently, the class of recurrent neural networks (\eg, LSTM) and graph neural networks are among the most popular architectures in \npas \cite{alon2018code2vec,alon2018code2seq,fernandes2018structured}.

\begin{comment}
    \input{motivating-example/boolean-exchange.tex}
    \input{motivating-example/loop-exchange.tex}
    \input{motivating-example/permute-stmt.tex}
    \input{motivating-example/switch-to-if.tex}
    \input{motivating-example/trycatch-insertion.tex}
    \input{motivating-example/unused-stmt-insertion.tex}
\end{comment}

\section{Evaluation Approach}
\label{sec:approach}

Our approach for evaluating \npas relies on a metamorphic relation that states: 
{\em the outputs of a \npa should not differ on semantically-equivalent programs}.
To this end, the evaluation approach is divided into two main steps: 
(1) generating new programs using semantic-preserving transformations, 
and (2) comparing the outputs of a \npa before and after the transformations to compute generalizability metrics.
We describe these steps in the rest of this section.

\subsection{Target Downstream Task}
We use the method name prediction task \mbox{\cite{allamanis2015suggesting,allamanis2016summarization}} in this work to evaluate the generalizability of \mbox{\npas}.
The goal of the task is to predict the name of a method given the body of the method.
This task has several applications such as code search \mbox{\cite{liu2019learning}}, code summarization \mbox{\cite{allamanis2016summarization}}, and code analogies \mbox{\cite{alon2018code2vec}}. 
Figure~\mbox{\ref{fig:motivating-example}} depicts an example of this task wherein \mbox{\npas} are given a method body and return candidate names for the method body, i.e., \texttt{compareTo} and \texttt{getCount}.
This task has been used as the downstream task to evaluate several state-of-the-art \mbox{\npas} \mbox{\cite{allamanis2017learning,alon2018code2vec,alon2018code2seq}}.

\subsection{Transformations}
In this work, we only evaluate \npas that take a method body as their input, therefore, 
we use the following set of transformations that are applicable to method-level code
to generate semantically-equivalent methods.
This set includes transformations ranging from common refactoring like variable renaming to more intrusive ones like loop exchange.
The goal is to evaluate the generalizability of \npas under a wide range of semantic-preserving changes to the structure of a method.  

\begin{itemize}[leftmargin=*]
    \item  \textbf{\VN (VN)} is a refactoring that renames the name of a variable in a method. The new name of the variable will be in the form of \texttt{varN} for a value of N such that \texttt{N} has not been defined in the scope. VN is a widely-used refactoring for methods.
    \item \textbf{\PS (PS)} swaps two independent statements (i.e., with no data or control dependence) in a basic block of a method.
    \item \textbf{\UN (UN)} inserts an unused string declaration to a randomly selected basic block in a method. Unused variables in methods are a common malpractice by developers. 
    \item \textbf{\LX (LX)} replaces \texttt{for} loops with \texttt{while} loops or vice versa.
    \item \textbf{\SF (SF)} replaces a \texttt{switch} statement in a method with an equivalent \texttt{if} statement.
    \item \textbf{\BX (BX)} switches the value of a boolean variable in a method from \texttt{true} to \texttt{false} or vice versa, and propagates this change in the method to ensure a semantic equivalence of the transformed method with the original method.
\end{itemize}

Note that each transformation has different impact on the structure of methods as follows. 

\begin{itemize}[leftmargin=*]
    \item The \VN transformation only changes the terminal values and does not affect the structure of an AST.
    \item The \PS transformation does not change the nodes, rather it only reorders two subtrees in an AST.
    \item The \UN transformation adds a few nodes into an AST, which increases the number of paths in the AST.
    \item The \LX transformation extensively impacts an AST by removing and inserting nodes.
    \item The \SF transformation also impacts the AST of a method substantially by removing and inserting nodes.
    \item The \BX transformation alters the value of \texttt{true} or \texttt{false} and modifies the structure of an AST by removing or inserting unary-not nodes.
\end{itemize}

\begin{comment}

Table~\ref{tab:types_of_trans} summarizes the type of code transformations that we used in our evaluations.

\begin{table}
    \centering
    \resizebox{\columnwidth}{!}{%
    \begin{tabular}{|@{\makebox[3em]{\rownumber\space}}|cZ|c|c|}
    \hline
        \gdef\rownumber{\stepcounter{magicrownumbers}\arabic{magicrownumbers}}

     Transformation & Kind & Change & Place \\ 
    \hline
    \hline

    Variable Renaming & ? & Rename the name of a variable & Single, All \\ \hline
    Loop Exchange & ? & $for$ loop with $while$ loop (and vice versa) & Single, All \\ \hline
    Switch to If & ? & $switch$ cases by $if-else$ conditions & Single, All \\ \hline
    Boolean Exchange & ? & $true$ with $false$ (and vice versa) & Single, All \\ \hline
    Permute Statement & ? & swap two independent statements & Pair of statements \\ \hline
    Try Catch Insertion & ? & Surround an expression with $try-catch$ & Random statement \\ \hline
    Unused Statement Insertion & ? & Insert an unused string declaration & Random position \\ \hline
    %Unreachable Statement Insertion & ? & Insert an $if$ block of $false$ condition & Random position \\ \hline
    \end{tabular}%
    }
    \caption{Types of transformations}
    \label{tab:types_of_trans}
\end{table}

\end{comment}

\subsection{Generalizability Metrics}
\label{sec:metrics}
In this study, we define a few metrics to measure different results of a \npa for transformed programs and thus to quantify the generalizability of the \npa.

Specifically, suppose $M$ denotes a set of methods, given a semantic-preserving program transformation $T$ that takes a method and creates a set $M'=\bigcup_{m\in M} T(m)$ of transformed methods, and a \npa $NPM:M\rightarrow L$, 
where $L$ denotes a set of labels, maps methods to labels.
We evaluate the generalizability of $NPM$ with respect to the transformation $T$, by comparing $NPM(m)$ and $NPM(m')$ for $m'\in T(m)$ for $m\in M$.
Ideally, the \npa should produce the same results on both $m$ and $m'$, that is $NPM(m)=NPM(m')$.
We define the following metrics. 
\begin{description}[leftmargin=0em] 
\item[Prediction Change Percentage.] We compute the prediction change percentage as follows:
\begin{equation}
	 PCP=\frac{|\{m' \in M'| NPM(m) \neq NPM(m')\}|}{|\{m' \in M'\}|}*100.
\end{equation}
The lower values of PCP for $NPM$ would suggest higher a degree of its generalizability with respect to the transformation. 

\item[Types of Changes.] Considering that the correctness of predicted labels of the $NPM$, five types of changes can happen:
\begin{enumerate}[(1),leftmargin=2em]
\item a correct prediction remains correct after the transformation,
\item a correct prediction changes to a wrong prediction after the transformation,
\item a wrong predicted label remains the same wrong label after the transformation,
\item a wrong prediction changes to a correct prediction after the transformation,
\item a wrong predicted label changes into a different, yet still wrong label after the transformation.
\end{enumerate}
We use the following five metrics to denote the proportion of each of these cases in the experiments. 
{\bf CCP, CWP, WWSP, WCP}, and {\bf WWDP} respectively denote the percentage of correct predictions that stay correct, the percentage of correct predictions that become wrong, the percentage of wrong predictions that stay to the same wrong prediction after the transformation, the percentage of wrong predictions that become correct, and the percentage of wrong predictions that change to a different wrong prediction after the transformation.

\item[Precision, Recall, and F$_{1}$-Score.]
We also use the traditional sub-token metrics (precision, recall and $F_1$-score) as commonly used in the literature for the method name prediction task~\mbox{\cite{alon2018code2vec, alon2018code2seq}} in this generalizability study. Suppose, $tp$ denotes the number of true positive sub-tokens, $fp$ denotes the number of false positive sub-tokens, and $fn$ denotes the number of false negative sub-tokens in the predicted method names.

\begin{itemize}
\item \mbox{\textit{Precision}} indicates the percentage of predicted sub-tokens that are true positives. It is the ratio of the correctly predicted positive sub-tokens to the total number of predicted positive sub-tokens:
$Precision = \frac{tp}{tp\,+\,fp}$

\item \mbox{\textit{Recall}} indicates the percentage of true positive sub-tokens that are correctly predicted. It is the ratio of the correctly predicted positive sub-tokens to the total number of sub-tokens in actual method names:
$Recall = \frac{tp}{tp\,+\,fn}$

\item \mbox{\textit{F1-Score}} is the harmonic mean of precision (P) and recall (R):

$F_{1}\text{--}Score = \frac{2}{P^{-1} + R^{-1}} = 2\,.\,\frac{P\,.\,R}{P\,+\,R}$
\end{itemize}

For example, a predicted name {\tt result\_compute} has two sub-tokens {\tt result} and {\tt compute}, and is considered as an exact match of the ground-truth name {\tt computeResult} which also has the same two sub-tokens (ignoring the case and the ordering of the tokens).
Similarly, a predicted name {\tt compute} has 100\% precision but only 50\% recall with respect to the same ground truth, and {\tt compute\_model\_result} has 100\% recall but only 67\% precision.
\end{description}

\section{Experimental Setting}
\label{sec:settings}

%\subsection{Models and Data}
%In this section, we describe the models\Fix{models or NPA?} and datasets that we used in the evaluation.

\subsection{Subject \NPAS}
The task of method name prediction \cite{allamanis2016summarization} has attracted some attention recently. We use three \npas that use different code representations and neural network characteristics for the task:
\ctv~\cite{alon2018code2vec}, \cts~\cite{alon2018code2seq}, and \ggnn~\cite{fernandes2018structured}.

\ctv~\cite{alon2018code2vec} uses a bag of AST paths to model source code.
Each path consists of a pair of terminal nodes and the corresponding path between them in the AST.
Each path, along with source and destination terminals, is mapped into its vector embeddings which are learned jointly with other network parameters during training.
The separate vectors of each path-context are then concatenated to a single context vector using a fully connected layer which is learned during training with the network.
An attention vector is also learned with the network; it is used to score each path-context and aggregate multiple path-contexts to a single code vector representing a method body.
After that, the model predicts the probability of each target method name given the code vector of the method body via a softmax-normalization between the code vector and each of the embeddings of all possible target method names.

While \ctv uses monolithic path embeddings and only generates a single label at a time, the \cts~\cite{alon2018code2seq} model uses an encoder-decoder architecture to encode paths node-by-node and generate labels as sequences at each step.
In \cts, the encoder represents a method body as a set of AST paths where each path is compressed to a fixed-length vector using a bi-directional LSTM which encodes paths node-by-node.
The decoder uses attentions to select relevant paths while decoding, and predicts sub-tokens of a target sequence at each step when generating the method name.

In \ggnn~\cite{fernandes2018structured}, a variety of semantic edges are added into the AST of a method body to construct a graph,
and the Gated Graph Neural Network (GGNN) is applied to encode such graphs~\cite{allamanis2017learning}.
The initial embedding for a node of the graph is the concatenation between the node type embedding and node token embedding. Then a fixed number of message passing steps are applied for a node to aggregate the embeddings of its neighbors.
The output of the GGNN encoder is then fed into a bi-directional LSTM decoder to generate the method name as a language model of sub-tokens~\cite{fernandes2018structured}.

\subsection{Datasets}
\label{sec:dataset}

%The datasets published along with \ctv only contain the programs in a preprocessed format, but, for this study, we needed the raw Java files to perform the transformations.
We have used the \cts dataset for training \npas for the study.
There are three Java datasets based on the GitHub projects: \JS, \JM, and \JL.

\begin{itemize}
\item \JS: This dataset contains $9$ Java projects for training, $1$ for validation and $1$ for testing. Overall, it contains about $700$K methods. The compressed size is about $366$MB and the extracted size is about 1.9GB.
\item \JM: This dataset contains $800$ Java projects for training, $100$ for validation and $100$ for testing. Overall, it contains about $4$M methods. The compressed size is about $1.8$GB and the extracted size is about $9.3$GB.
\item \JL: This dataset contains $9000$ Java projects for training, $200$ for validation and $300$ for testing. Overall, it contains about $16$M methods. The compressed size is about $7.2$GB and the extracted size is about $37$GB.
\end{itemize}

\begin{table}[t]
    \begin{center}
        \caption{Performance of trained models for method name prediction in the testing dataset.}
        \def\arraystretch{1.1}
        \label{table:all_models}
        \resizebox{0.98\columnwidth}{!}{%
        \begin{tabular}{|c|c|R{4cm}|c|c|c|c|}
            \hline
            \textbf{Model} & \textbf{Dataset} & 
            \textbf{\# Original methods in the testing dataset} &
            \textbf{Precision} & \textbf{Recall} & \textbf{$F_1$-Score} \\ \hline 
            \hline
            \multirow{3}{*}{\ctv} & \JS &  44426 & 28.36 & 22.37 & 25.01 \\ \cline{2-6}
                                  & \JM & 351628 & 42.55 & 30.85 & 35.76 \\ \cline{2-6}
                                  & \JL & 370930 & 45.17 & 32.28 & 37.65 \\ \hline
            \hline 
            \multirow{3}{*}{\cts} & \JS &  44426 & 46.30 & 38.81 & 42.23 \\ \cline{2-6}
				                  & \JM & 351628 & 59.94 & 48.03 & 53.33 \\ \cline{2-6}
				                  & \JL & 370930 & 64.03 & 55.02 & 59.19 \\ \hline
            \hline 
		    \multirow{3}{*}{\ggnn} & \JS &  44426 & 49.12 & 47.18 & 48.59 \\ \cline{2-6}
                                   & \JM & 351628 & 58.30 & 47.49 & 52.34 \\ \cline{2-6}
                                   & \JL & 370930 & 60.76 & 50.32 & 55.23 \\ \hline
        \end{tabular}%
        }
    \end{center}
\end{table}

\subsection{Training Models per Datasets}
The authors of \ctv and \cts have made the source code public for training and evaluating their models.
For \ggnn, the implementation of the network is available but the code graph generation is not; so we re-implement the step to generate graphs. We use the parser SrcSlice\footnote{https://github.com/srcML/srcSlice}, an extension of SrcML\footnote{https://www.srcml.org/, +400 node types for supporting multiple programming languages}, to produce data dependency edges among AST nodes for training \ggnn. 

We train each model for the method name prediction task with the configurations described in their original papers on each of the three aforementioned datasets, and thus construct three \ctv, three \cts, and three \ggnn{} \mbox{\npas}. 
Similar to the state-of-the-art approaches, i.e. \mbox{\cite{alon2018code2vec,alon2018code2seq}}, we train models on the training set, tune on the validation set for maximizing $F_1$-score, and finally report results on the unseen testing set.
Table~\mbox{\ref{table:all_models}} summarizes the performance of trained models for method name prediction on the testing set.
While the performance of our trained models for \cts is on par with to the ones reported in the corresponding paper~\cite{alon2018code2seq}, the performance of \ctv did not reach the performance reported in \cite{alon2018code2vec}, due to the differences in the dataset. 
However, the performance of our trained \ctv models is similar to the one reported in \cite{alon2018code2seq}.
For \ggnn, the performance is reasonably different from what were reported in \cite{fernandes2018structured} for mainly three reasons: 
(1) the ASTs produced by our parser are different,
(2) the extraction of some types of semantic edges proposed in \cite{fernandes2018structured} requires expensive analysis of the methods; therefore, we implemented and included only a subset (seven out of ten) of semantic edges into the ASTs when constructing the graphs, and 
(3) the datasets are different.

\subsection{Population of Transformed Programs}
We have used our own tool based on the \texttt{JavaParser}\footnote{https://github.com/javaparser/javaparser} library to transform Java methods. Henceforth we use terms program and method interchangeably.
Two authors were involved in the implementation, testing and code review. We have performed manual inspection of sample transformed programs to ensure correctness of the transformations. 

We have applied the applicable transformations to the methods available in the {\em testing} data of the three datasets mentioned in Section~\ref{sec:dataset}.
The number of original methods in our study is $1,415,116$,\footnote{This total number is different from the numbers in Table~\ref{table:all_models} because a method in the testing dataset may contain code elements eligible for multiple types of transformations and be counted multiple times.}. 
Overall, the number of original methods with incorrect predictions is, on average, $2.8$ times higher than the number of methods with correct predictions.

We create a set of \emph{single-place} transformed programs (Table~\ref{table:cop_summary}) by applying transformations to each eligible location in methods separately resulting in $2,822,810$ transformed methods; e.g., if a method has three eligible locations for a transformation, we would generate three distinct methods by transforming each individual location separately.    
The types and number of applicable transformations vary from a method to another.
Therefore, in our approach, different methods, based on the language features that they use, produce a different number of transformed programs.
In total, the number of transformed programs generated from the programs with incorrect initial predictions is much higher (4.2x and higher) than the number of transformed programs generated from the programs with correct initial predictions, which may suggest that programs with correct predictions may be smaller and simpler.

\paragraph{\textbf{Artifacts}} The source code of the program transformation tool and the datasets of the transformed programs used in this paper are publicly available at \url{https://github.com/mdrafiqulrabin/tnpa-generalizability/}.

\begin{comment}
    Moreover, programs with incorrect predictions are amenable to, on average, 1.8 times more transformations compared to programs with correct predictions, which may suggest that programs with correct predictions may be smaller and simpler. 
\end{comment}

\subsection{Research Questions}
In this paper, we seek to answer the following research questions.
\begin{itemize}
    \item[RQ1] How do the transformations impact the predictions of \npas in the single-place transformed dataset?
    \item[RQ2] When do the transformations affect \npas the most?
    \item[RQ3] How does the method length impact the generalizability of \npas?
    \item[RQ4] What are the trends in types of changes?
    \item[RQ5] How do the transformations affect the precision, recall and $F_1$-score of the \npas?
\end{itemize}

\section{Results}
\label{sec:results}

\subsection{RQ1: Impact of Transformations on the Predictions of \NPAS}
\begin{table*}
    \caption{Prediction Change Percentage (PCP) across all models, datasets, and transformations.}
    \label{table:cop_summary}
    \def\arraystretch{1.1}
    \resizebox{0.9\textwidth}{!}{%
    \begin{tabular}{|c|r|r|r|r|r|r|}
        \hline
        \multirow{2}{*}{\textbf{Transformation}}
        & \multirow{2}{*}{\textbf{Dataset}}
        & \multirow{2}{*}{\textbf{\# Original methods}}
        & \multirow{2}{*}{\textbf{\# Transformed methods}}
        & \multicolumn{3}{c|}{\textbf{Prediction change (\%) (PCP)}} \\ \cline{5-7}
        & & & & \textbf{\ctv} & \textbf{\cts} & \textbf{\ggnn} \\ 
        \hline \hline
        
        \multirow{4}{*}{\VN}
        & \JS & 31113 & 123123 & 54.92 & 57.16 & {\bf 28.17} \\  \cline{2-7}
        & \JM & 235961 & 771208 & 46.55 & 48.75 & {\bf 35.96} \\  \cline{2-7}
        & \JL & 252725 & 916565 & 42.06 & 47.04 & {\bf 31.92} \\  \cline{2-7}
        & \multicolumn{3}{r|}{Weighted Average = } & 44.85 & 48.46 & {\bf 33.39} \\  
        \hline\hline  
        
        \multirow{4}{*}{\BX}
        & \JS & 1158 & 1519 & 53.85 & 54.31 & {\bf 29.37} \\  \cline{2-7}
        & \JM & 6407 & 8840 & 50.35 & 44.71 & {\bf 33.74} \\  \cline{2-7}
        & \JL & 8868 & 12107 & 47.80 & 51.43 & {\bf 31.98} \\  \cline{2-7}
        & \multicolumn{3}{r|}{Weighted Average = } & 49.21 & 48.98 & {\bf 32.50} \\  
        \hline\hline  
        
        \multirow{4}{*}{\LX}
        & \JS & 3699 & 5160 & 59.38 & 52.54 & {\bf 31.66} \\  \cline{2-7}
        & \JM & 17107 & 23533 & 62.77 & 45.29 & {\bf 36.67} \\  \cline{2-7}
        & \JL & 35565 & 49665 & 46.52 & 42.51 & {\bf 31.75} \\  \cline{2-7}
        & \multicolumn{3}{r|}{Weighted Average = } & 52.25 & 44.01 & {\bf 33.22} \\  
        \hline\hline  
        
        \multirow{4}{*}{\SF}
        & \JS & 246 & 259 & 68.73 & 61.78 & {\bf 31.45} \\  \cline{2-7}
        & \JM & 3312 & 3839 & 59.91 & {\bf 41.60} & 43.73 \\  \cline{2-7}
        & \JL & 10478 & 11165 & 30.33 & {\bf 29.08} & 45.50 \\  \cline{2-7}
        & \multicolumn{3}{r|}{Weighted Average = } & 38.42 & {\bf 32.78} & 44.82 \\  
        \hline\hline  
        
        \multirow{4}{*}{\PS}
        & \JS & 3397 & 9169 & 72.80 & 57.32 & {\bf 26.36} \\  \cline{2-7}
        & \JM & 16150 & 44711 & 65.44 & 42.64 & {\bf 34.09} \\  \cline{2-7}
        & \JL & 21956 & 74973 & 64.38 & 41.93 & {\bf 26.32} \\  \cline{2-7}
        & \multicolumn{3}{r|}{Weighted Average = } & 65.35 & 43.27 & {\bf 29.02} \\  
        \hline\hline  
        
        \multirow{4}{*}{\UN}
        & \JS & 44426 & 44426 & 39.97 & 45.60 & {\bf 28.34} \\  \cline{2-7}
        & \JM & 351621 & 351621 & {\bf 35.80} & 40.25 & 42.79 \\  \cline{2-7}
        & \JL & 370927 & 370927 & {\bf 31.21} & 37.44 & 35.67 \\  \cline{2-7}
        & \multicolumn{3}{r|}{Weighted Average = } & {\bf 33.82} & 39.20 & 38.51 \\  
        \hline 
             
        \end{tabular}%
        }
\end{table*}

Table~\ref{table:cop_summary} shows the prediction change percentage (PCP) of the \npas for each transformation and dataset.
In this table, ``\# Original methods'' denotes the number of methods eligible for the corresponding transformation, ``\# Transformed methods'' denotes the number of methods generated as the result of applying the corresponding transformations on the original methods, and ``Prediction change (\%)'' denotes PCP as defined in Section~\ref{sec:approach}.
``Weighted Average'' provides the weighted average of PCP for each transformation and neural models.
The bold values in the Table~\ref{table:cop_summary} highlight the minimum value of PCP for the transformations.  
Since a transformation can be applied in more than one place separately in a method body, the number of transformed methods can be larger than the number of original methods. 

As Table~\ref{table:cop_summary} depicts, all \npas are likely to susceptible to semantic-equivalent transformations; however, the impact of transformations on PCP differs among different neural networks and datasets. Overall, \ggnn seems less prone to prediction changes; in 14 out of 18 cases, PCP in \ggnn is significantly less than \ctv and \cts. Moreover, in four out of six transformations, the weighted average of PCP for \ggnn is lower than the rest.

\observation{In most cases, \ggnn seems less susceptible to prediction changes under semantic-preserving transformations, compared to \ctv and \cts.}

Within \ctv, \cts, and \ggnn, the PCP trend varies for different transformations and datasets. 
\ctv is comparatively most sensitive to \PS on all datasets. On the other hand, \cts is most vulnerable to \SF in \JS, \VN in \JM, and \BX in \JL. In \ggnn, \SF is the most powerful transformation on all datasets.
In most cases, for \ctv and \cts, the PCP for \UN is comparatively less than the other transformations, except for \cts in \JL where \SF is less sensitive. In \ggnn, \PS is a comparatively less powerful transformation than others on all datasets.
Overall, based on the weighted average, it is likely that, \ctv is most sensitive to \PS and least sensitive to \UN, \cts is most sensitive to \BX and least sensitive to \SF, and \ggnn is most sensitive to \SF and least sensitive to \PS.

Based on the weighted average, \ggnn performs worst for \SF and \UN transformations. These two transformations add some additional nodes and paths in the AST. For \ctv and \cts, if models give less attention to those new paths, then the change can less effective. However, \ggnn works by using a message passing mechanism among the nodes with a limited number of passing steps. In \UN, because there is some irrelevant information added into the code, the passing steps in \ggnn can capture this information and ignore other useful information, thus having a strong impact on the prediction results.
In \SF, because the structure of the AST is modified by adding and removing nodes, and \ggnn is a node-based method, i.e., combining node information with message passing, thus the \ggnn can sensitive to node modification in the AST for \SF.

Table~\ref{table:cop_summary} also supports that, in most cases, \PS is more powerful than \VN in \ctv model whereas \VN is more powerful than \PS in \cts model. This is probably caused by the real-value embeddings of AST paths are different for \ctv and \cts. In \ctv, an embedding matrix is initialized randomly for paths and learned during training, that contains rows that are mapped to each of the AST paths. On the other hand, in \cts, each node of a path comes from a learned embedding matrix, and then a bi-directional LSTM is used to encode each of the AST paths separately. The bi-directional LSTM reads the path once from beginning to the end (as original order) and once from end to beginning (in reverse order). Therefore, the order changes by \PS may become less sensitive to \cts than \ctv.

Another observation is that, in most cases of \ctv and \cts, the PCP of the transformations in \JS is high, and it is significantly lower on larger datasets, \ie, \JM, and \JL. 
In \ggnn, the PCP of the transformations shows a different trend: lowest in \JS, in most cases, and highest in \JM.

\observation{In most cases, the effect of prediction change for \ctv and \cts is reduced as the dataset size increases, compared to \ggnn.}

\subsection{RQ2: When Transformations Affect \NPAS the Most?}
\begin{figure*}
\noindent \begin{minipage}{.33\textwidth}
\caption*{(a) \ctv}
\includegraphics[width=0.99\linewidth]{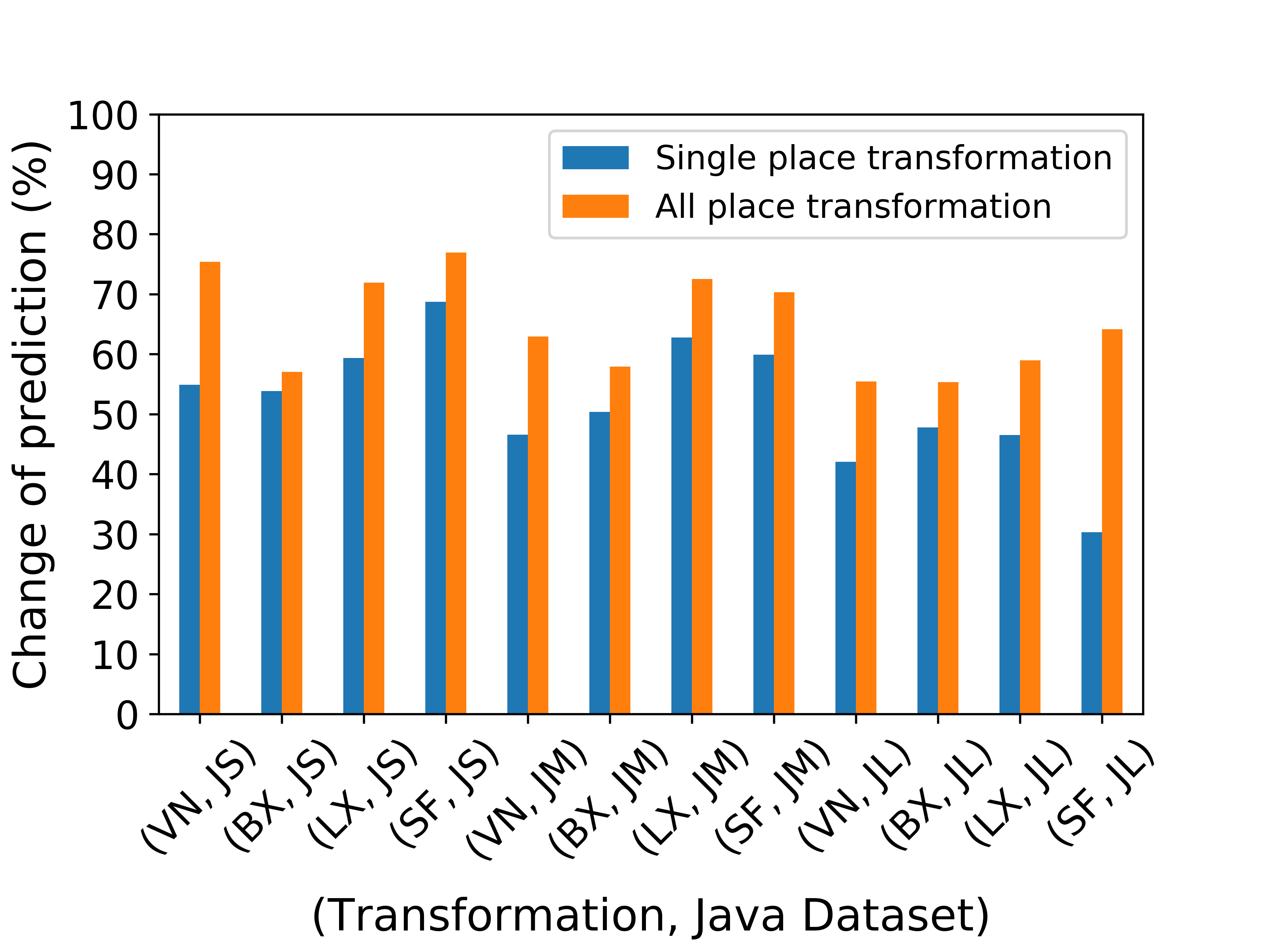}
\end{minipage}%
\begin{minipage}{.33\textwidth}
\caption*{(b) \cts}
\includegraphics[width=0.99\linewidth]{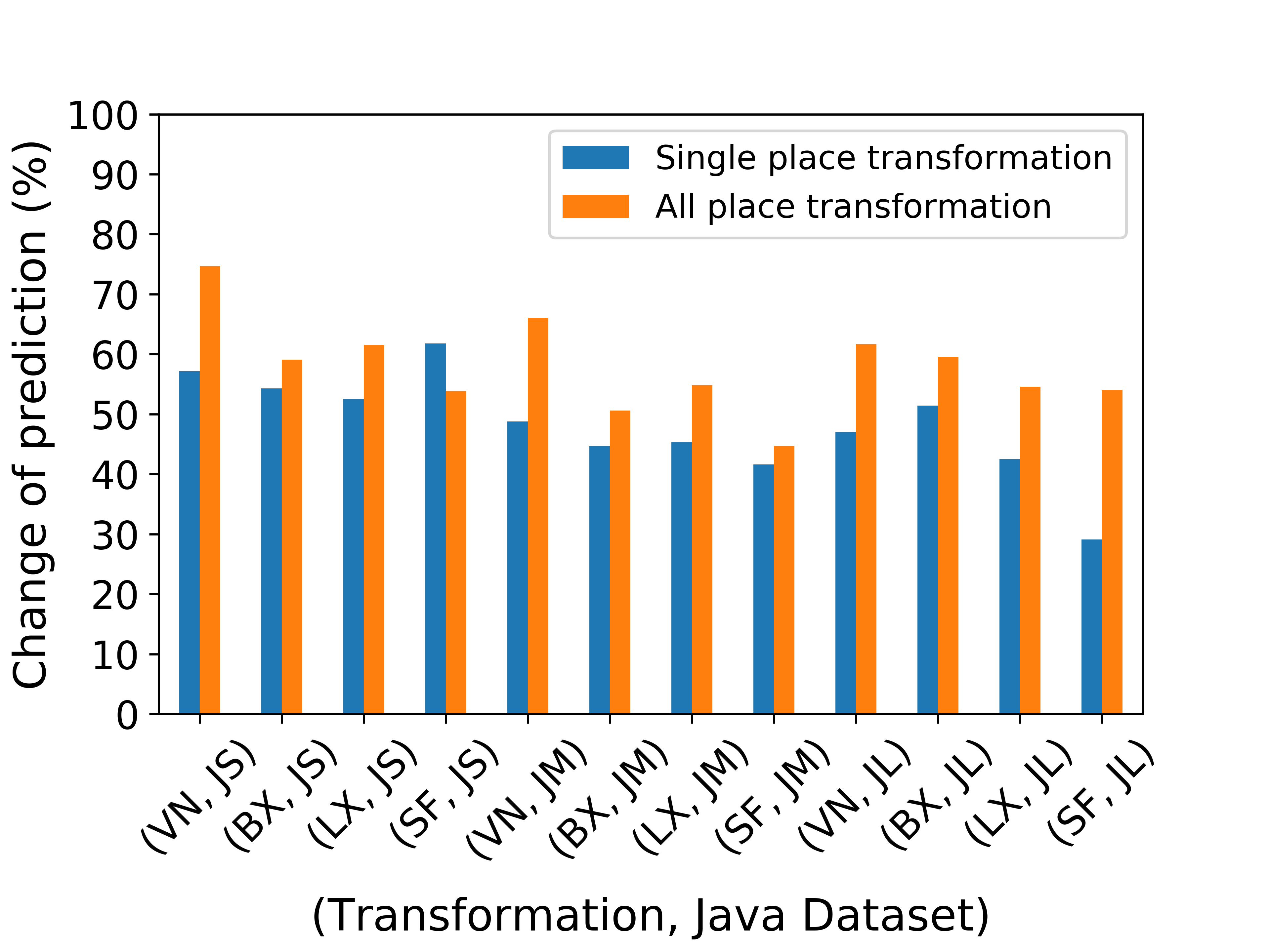}
\end{minipage}%
\begin{minipage}{.33\textwidth}
\caption*{(c) \ggnn}
\includegraphics[width=0.99\linewidth]{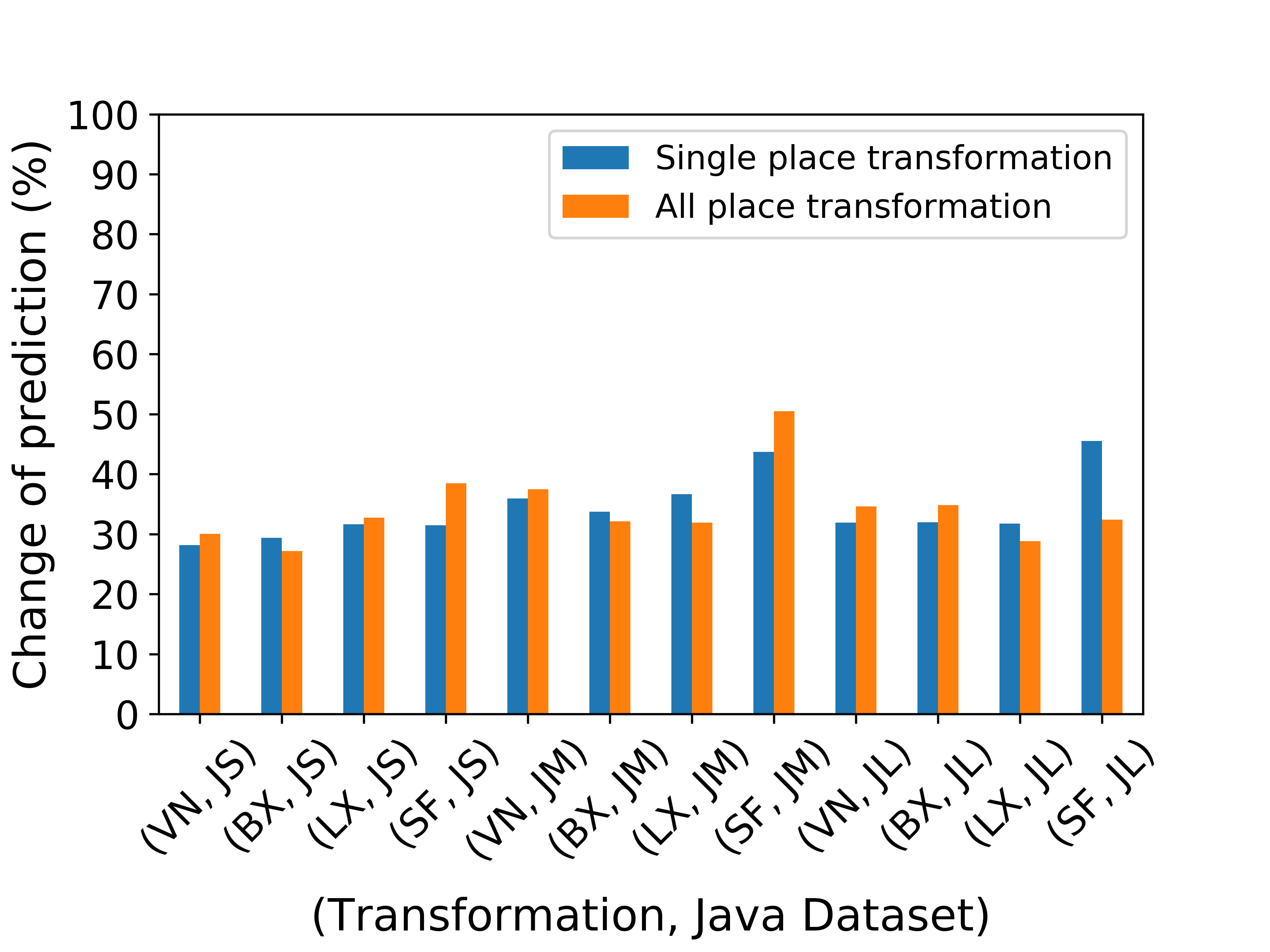}
\end{minipage}
\caption{Prediction Change Percentage (PCP) in single-place vs. all-place transformations.}
\label{fig:single_vs_all}
\end{figure*}

\begin{figure*}
\noindent \begin{minipage}{.33\textwidth}
\caption*{(a) \ctv}
\includegraphics[width=0.99\linewidth]{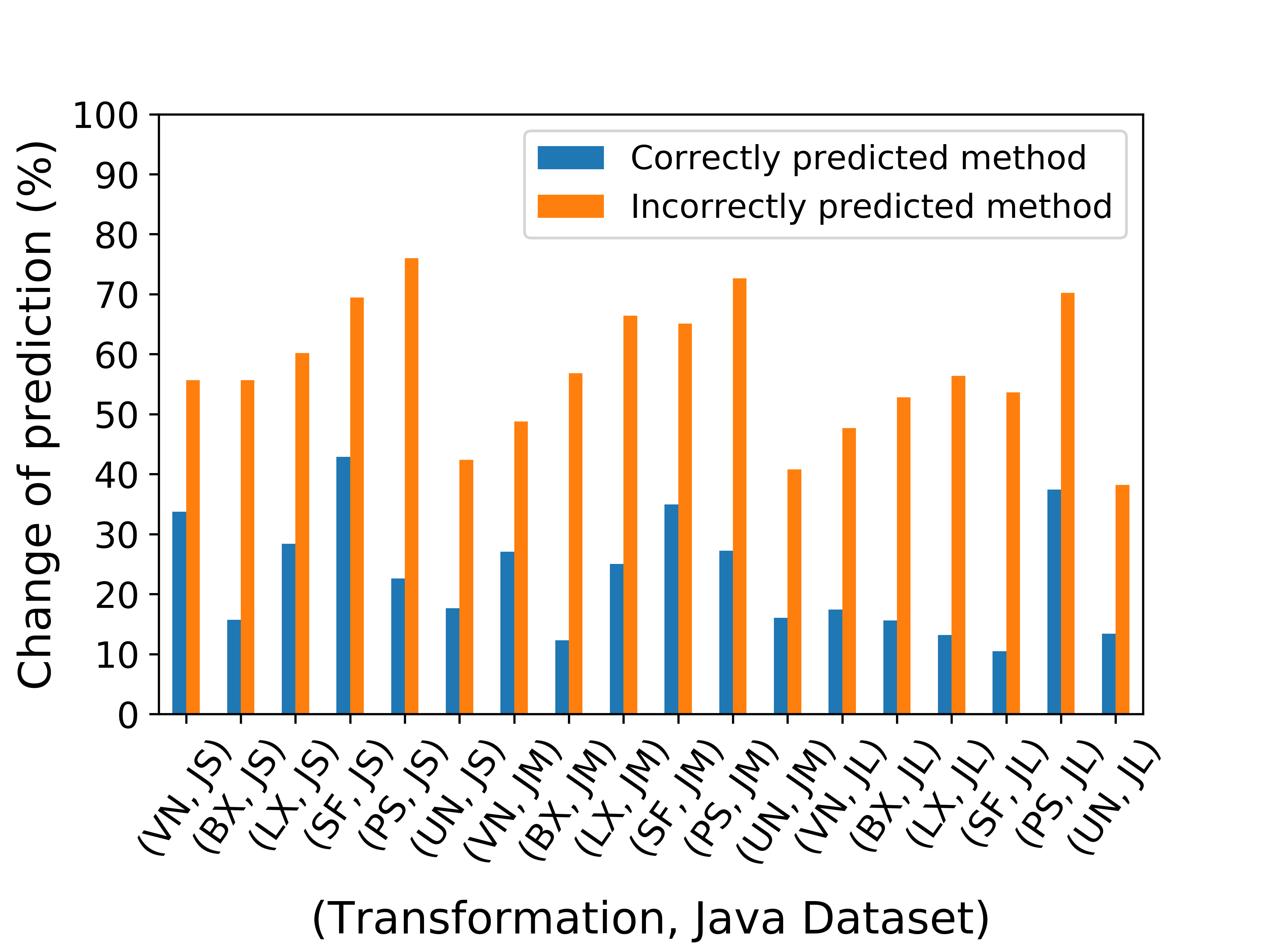}
\end{minipage}%
\begin{minipage}{.33\textwidth}
\caption*{(b) \cts}
\includegraphics[width=0.99\linewidth]{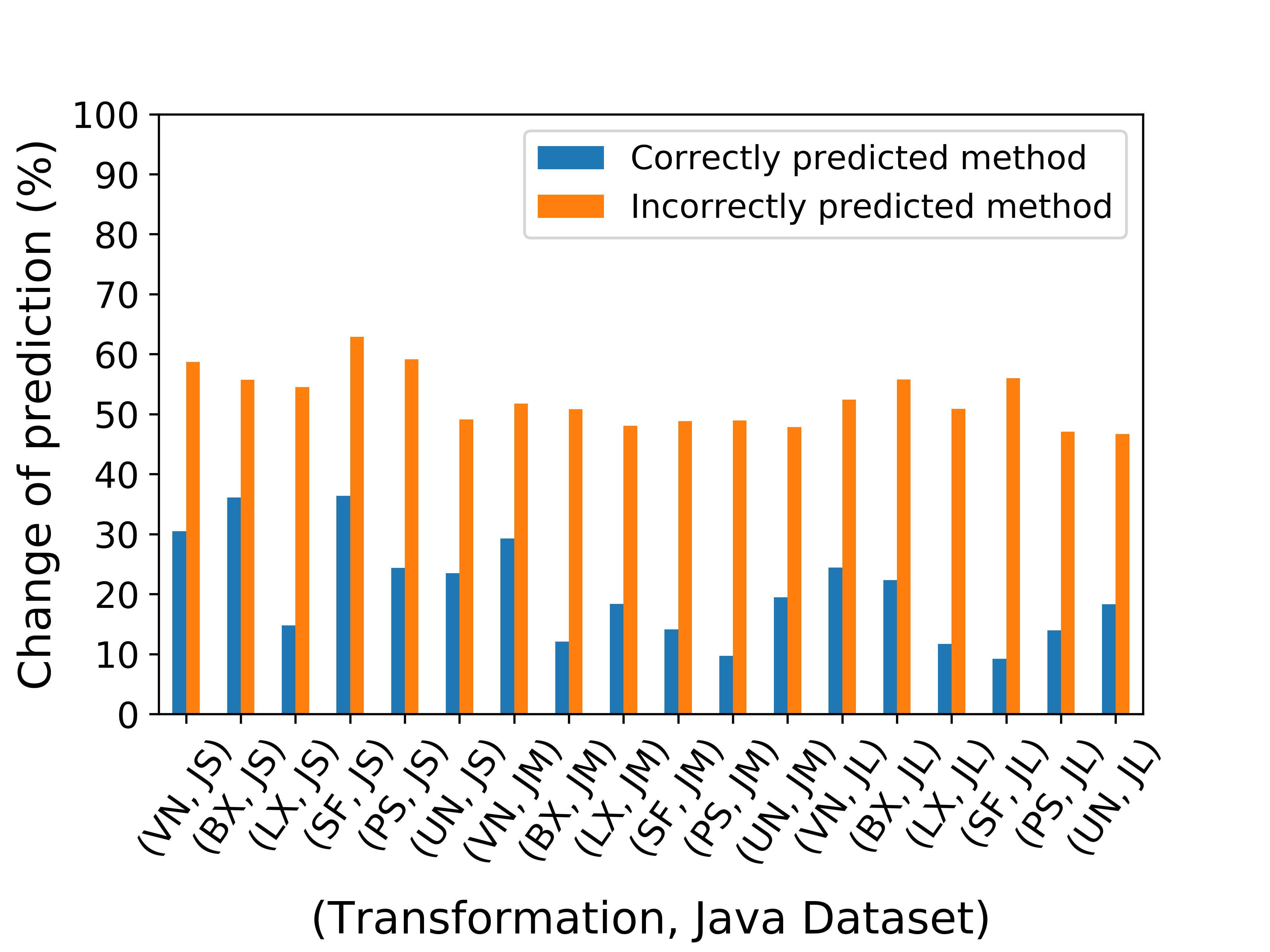}
\end{minipage}%
\begin{minipage}{.33\textwidth}
\caption*{(c) \ggnn}
\includegraphics[width=0.99\linewidth]{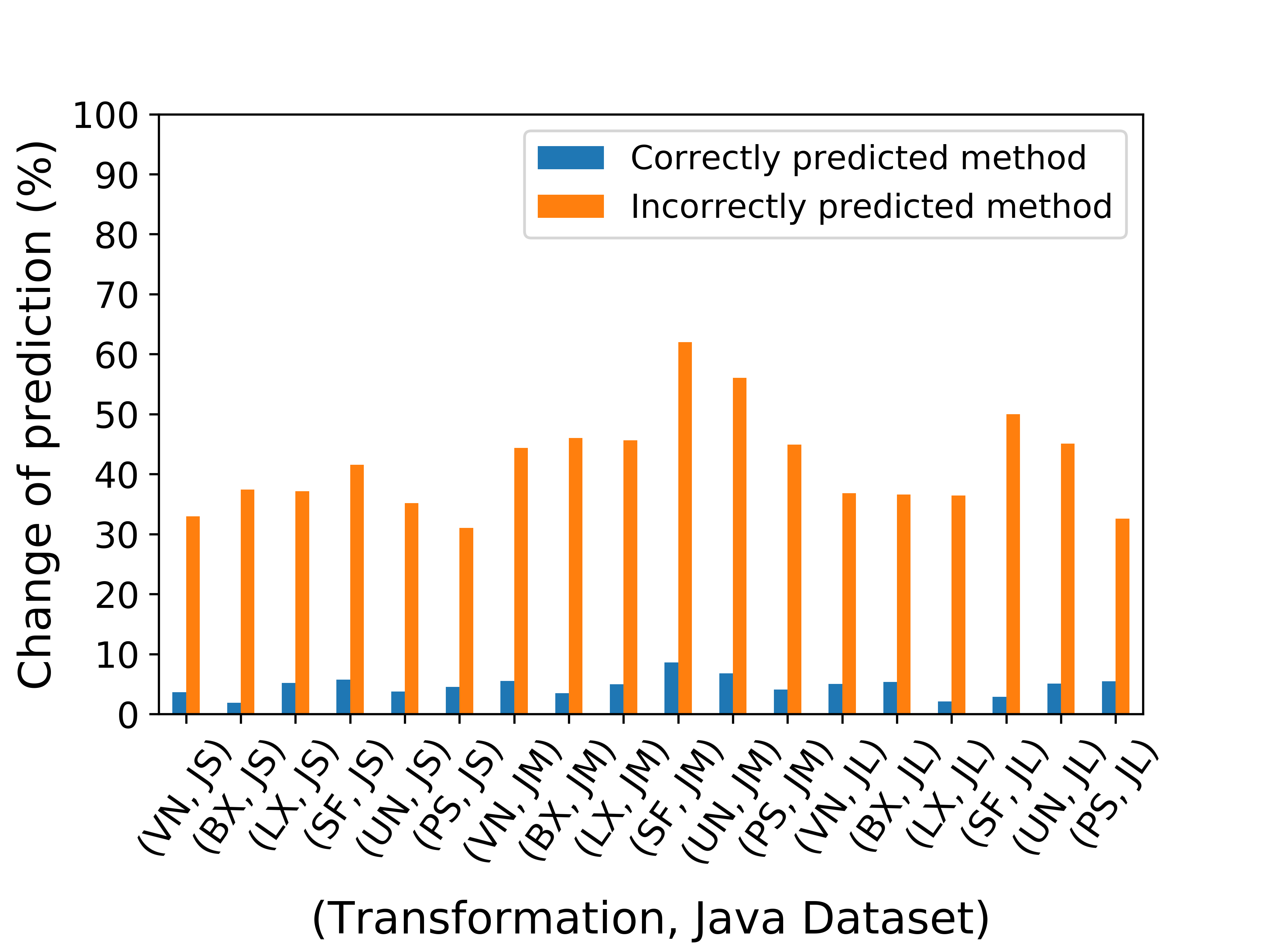}
\end{minipage}
\caption{Prediction Change Percentage (PCP) on correctly vs. incorrectly predicted methods.}
\label{fig:correct_vs_incorrect}
\end{figure*}

\subsubsection{Single-place transformation vs. All-place transformation}

In our analysis, thus far, if a program has multiple candidates for a transformation, say $n$ candidates, for transformation, we only apply them one at the time and end up with $n$ distinct transformed programs. We call this \emph{single-place} transformation.
Alternatively, we can apply the transformations to \emph{all} candidate locations in the program simultaneously to create only one transformed program. 
We call this \emph{all-place} transformation.
We evaluate the generalizability of \npas under all-place transformation for the following transformations: \VN, \BX, \LX, and \SF.
Note that the all-place transformation is not applicable to \PS and \UN transformations, as we apply the \PS on a pair of statements and the \UN on a random block.

Figure~\ref{fig:single_vs_all} compares the impact of single-place transformation and all-place transformation on the prediction changes in all \npas.
For the \ctv model, the percentage of prediction change for the all-place transformation is higher than the single-place transformation by a good margin for all the cases. 
Similarly, for the \cts model, the percentage of prediction change for the all-place transformation is higher than the single-place transformation by a good margin except for the case (\SF, \JS).
After a closer examination of \JS dataset and \SF transformation, we observe that the number of transformed methods for all-place is only $13$, which is too low to provide comparative insight.
For the \ggnn model, the difference between all-place transformation and single-place transformation is relatively very small compared to the \ctv and \cts models. 
Even for (\BX, \JS + \JM), (\LX, \JM + \JL), and (\SF,\JL), the percentage of prediction changes for the single-place transformation is higher than the all-place transformation. The results may suggest that the performance of \ggnn under single-place transformations and all-place transformations is almost consistent.

\observation{While all-place transformations are more likely to induce prediction changes in \ctv and \cts than single-place transformations, the performance of \ggnn remains relatively similar under both types of transformations.}

\subsubsection{Correctly predicted methods vs. Incorrectly predicted methods}

We also evaluate the generalizability of \npas under correctly and incorrectly predicted methods.
Figure~\ref{fig:correct_vs_incorrect} compares the impact of correctly predicted methods and incorrectly predicted methods on the prediction changes in all \npas.
In the \ctv model, the percentage of changes in predictions after transformation in the correctly predicted methods ranges from 10.45\% to 42.86\%, while, in the incorrectly predicted methods, a larger portion of transformations, 38.18\% to 76.00\%, change the prediction of \ctv.
Similarly, in the \cts model, the percentage of changes in predictions after transformation in the correctly predicted methods and in the incorrectly predicted methods ranges from 9.19\% to 36.36\% and 46.66\% to 62.90\%,  respectively.
However, in the \ggnn model, while the percentage of changes in predictions after transformation on the correctly predicted methods ranges from 1.90\% to 8.58\%, the percentages range from 31.05\% to 62.01\% in the incorrectly predicted methods.

\observation{It is likely that \ggnn is more stable than \ctv and \cts in the originally correct methods, and the changes in prediction happen more frequently in the originally incorrect methods for all models.}

\subsubsection{The Effect of \texorpdfstring{$X\%$-Transformation}{}}

%\begin{landscape}
\begin{table*}
    \caption{The PCP for $X\%$-transformations across different datasets and models.}
    \label{table:x_percent}
    \def\arraystretch{1.25}
    \resizebox{\textwidth}{!}{%
    \begin{tabular}{|c|c|r|c|c|c|c|c|c|c|c|c|}
        \hline
        \multirow{2}{*}{\textbf{Dataset}} 
        & \multirow{2}{*}{\textbf{Transformation}} 
        & \multirow{2}{*}{\textbf{\pbox{10cm}{\# Transformed \\ methods}}}
        & \multicolumn{3}{c|}{\textbf{25\% Transformation}} 
        & \multicolumn{3}{c|}{\textbf{50\% Transformation}} 
        & \multicolumn{3}{c|}{\textbf{75\% Transformation}}  \\ \cline{4-12}
        & & & \ctv & \cts & \ggnn & \ctv & \cts & \ggnn & \ctv & \cts & \ggnn
        \\ \hline \hline
        
        \multirow{4}{*}{\JS}
        & \VN & 15937 & 63.29 & 54.36 & {\bf 29.56} & 71.88 & 65.89 & {\bf 29.87} & 75.18 & 70.57 & {\bf 30.36} \\ \cline{2-12}
        & \BX & 75 & 80.00 & 63.00 & {\bf 37.70} & 79.67 & 64.67 & {\bf 36.07} & 79.66 & 64.66 & {\bf 32.79} \\ \cline{2-12}
        & \LX & 302 & 81.95 & 65.90 & {\bf 32.44} & 81.87 & 65.98 & {\bf 32.44} & 81.38 & 65.56 & {\bf 34.73} \\ \cline{2-12}
        & \SF & 0 & - & - & - & - & - & - & - & - & - \\ \hline
        \hline 
        
        \multirow{4}{*}{\JM}
        & \VN & 101003 & 54.07 & 46.51 & {\bf 37.18} & 62.65 & 57.77 & {\bf 38.91} & 66.51 & 63.20 & {\bf 37.88} \\ \cline{2-12}
        & \BX & 428 & 69.66 & 48.20 & {\bf 31.37} & 70.50 & 48.60 & {\bf 30.97} & 70.91 & 47.49 & {\bf 31.17} \\ \cline{2-12}
        & \LX & 1292 & 86.01 & 55.11 & {\bf 28.72} & 86.11 & 57.37 & {\bf 26.29} & 85.37 & 57.62 & {\bf 29.01} \\ \cline{2-12}
        & \SF & 98 & 82.91 & 43.62 & {\bf 55.42} & 84.44 & 44.90 & {\bf 50.00} & 89.03 & 45.16 & {\bf 57.08} \\ \hline
        \hline 
        
        \multirow{4}{*}{\JL}
        & \VN & 114748 & 45.62 & 43.23 & {\bf 33.37} & 53.32 & 53.99 & {\bf 35.75} & 56.98 & 58.83 & {\bf 34.59} \\ \cline{2-12}
        & \BX & 642 & 71.81 & 59.03 & {\bf 28.18} & 71.09 & 62.76 & {\bf 31.76} & 71.93 & 62.03 & {\bf 30.40} \\ \cline{2-12}
        & \LX & 2899 & 79.77 & 56.00 & {\bf 25.96} & 79.02 & 56.79 & {\bf 27.41} & 78.57 & 56.92 & {\bf 27.05} \\ \cline{2-12}
        & \SF & 125 & 69.00 & 56.00 & {\bf 34.78} & 73.60 & 56.40 & {\bf 32.36} & 73.60 & 56.60 & {\bf 34.95} \\ \hline

        \end{tabular}%
        }
\end{table*}
%\end{landscape}

In this section, we evaluate the generalizability of \mbox{\npas} under $X\%$-transformation for the following transformations: \mbox{\VN}, \mbox{\BX}, \mbox{\LX}, and \mbox{\SF}.
%Note that we apply the \mbox{\PS} on a pair of independent statements and the \mbox{\UN} on one randomly selected block in a method, and thus $X\%$-transformation is not suitable for them.
If a transformation $t$ is applicable to $n$ locations in a method body, $X\%$-transformation randomly picks $\floor{\frac{n*X}{100}}$ of those locations and applies $t$ to create a new transformed program. The number of all $X\%$-transformed programs grows exponentially with the number of locations; therefore,
to manage the complexity, in $X\%$-transformation we randomly pick the locations in a method body, instead of considering all possible combinations, to create transformed programs. 
We study the $X\%$-transformation with $X=\{25, 50, 75\}$.
For each transformation $t$, we first create a dataset $d_t^4$ that contains methods with four or more possible locations, so that the transformation $t$ is applicable to each method for each $X=\{25, 50, 75\}$, e.g., methods with at least four variables for \mbox{\VN}.
To account for the randomness, we run each setting five times and report the average results, except for the \mbox{\ggnn} that we run only three times due to the longer graph construction and processing time.

Table \mbox{\ref{table:x_percent}} shows the results of the $X\%$-transformations.
In each $X\%$-transformation, \mbox{\ggnn} has a much lower PCP value than the \mbox{\ctv} and \mbox{\cts} models for all the transformations across the three Java datasets.
Moreover, in \mbox{\ggnn} models, the differences of PCP under different $X$ are relatively small (mostly a few percentage points) and do not yield a clear trend.
On the other hand, in \mbox{\ctv} and \mbox{\cts} models, with the \mbox{\VN} transformation, the PCP tends to increase as $X$ grows, but with other transformations expect \mbox{\VN}, the PCP shows modest changes only.
Note that in $X\%$-transformation, compared to \mbox{\VN}, the numbers of transformed programs for other transformations are much lower, which might be too low to provide statistical significance or comparative insights.

\observation{The performance of \mbox{\ggnn} in terms of PCP remains similar in all cases under $X\%$-transformation, but the PCP of \mbox{\ctv} and \mbox{\cts} for \mbox{\VN} increases as $X$ grows.}

\subsection{RQ3: Impact of Method Length on Generalizability}
\begin{figure*}
\noindent \begin{minipage}{.33\textwidth}
\caption*{(a) \ctv (\JS)}
\includegraphics[width=0.99\linewidth]{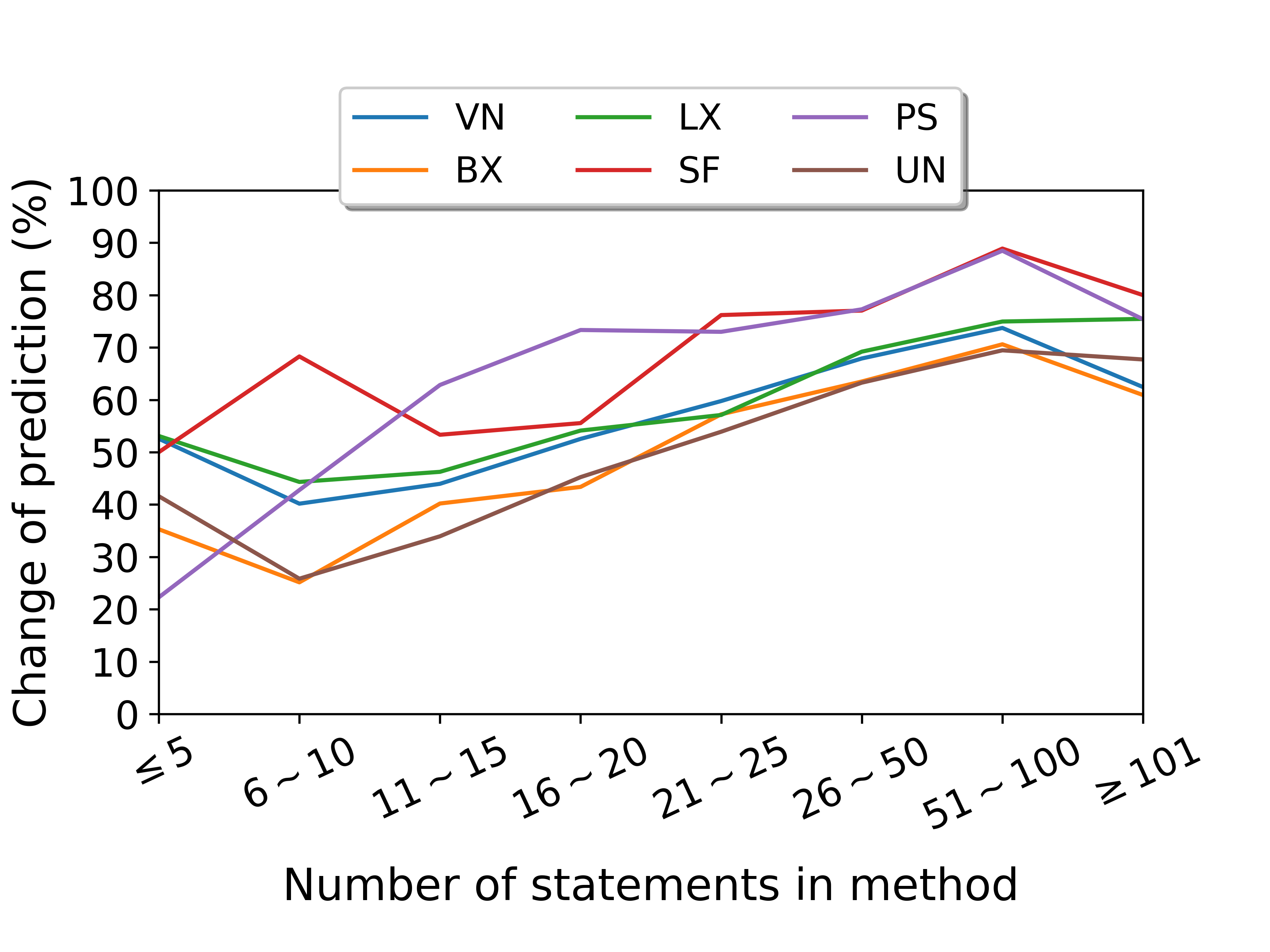}
\end{minipage}%
\begin{minipage}{.33\textwidth}
\caption*{(b) \ctv (\JM)}
\includegraphics[width=0.99\linewidth]{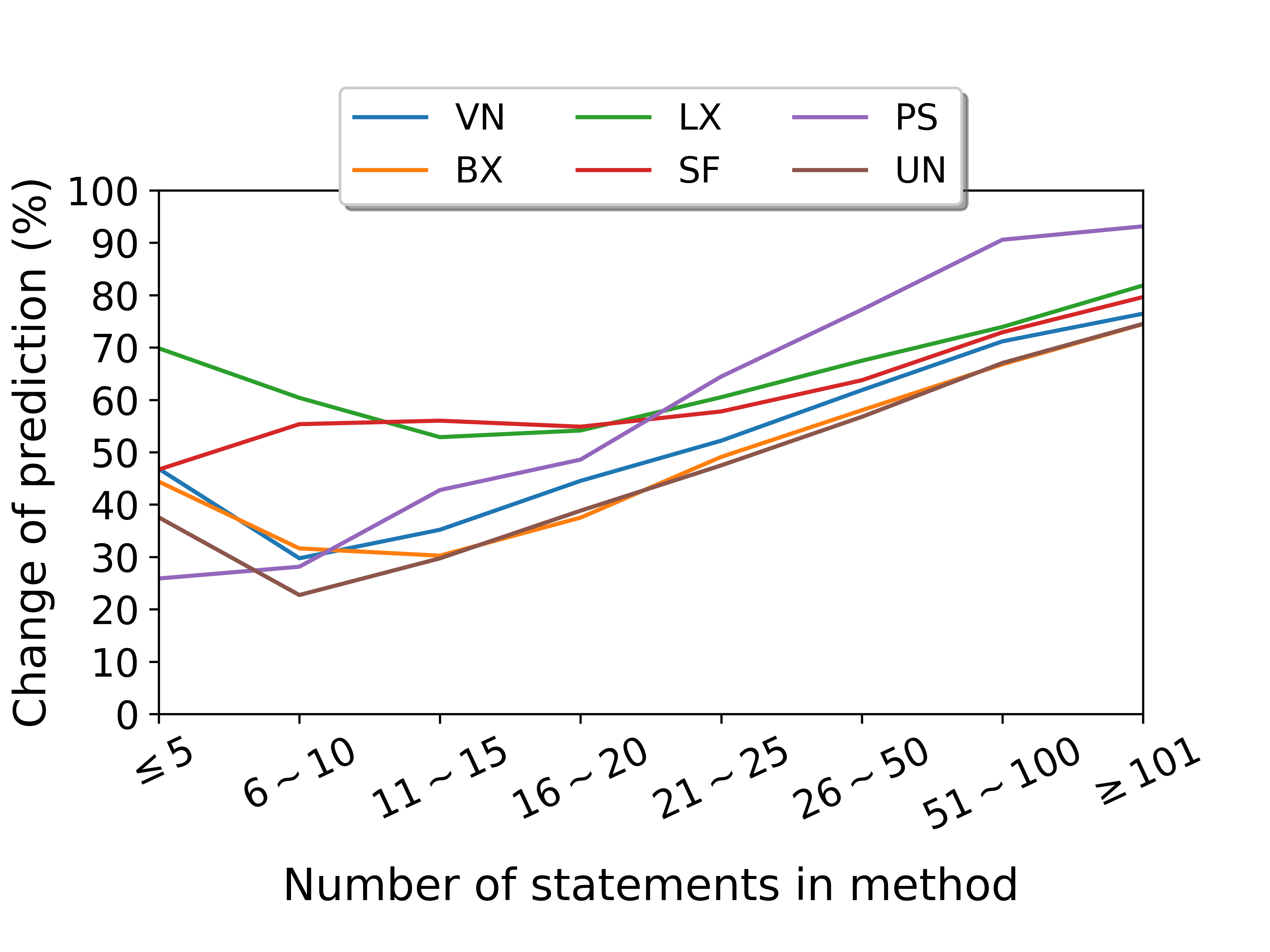}
\end{minipage}%
\begin{minipage}{.33\textwidth}
\caption*{(c) \ctv (\JL)}
\includegraphics[width=0.99\linewidth]{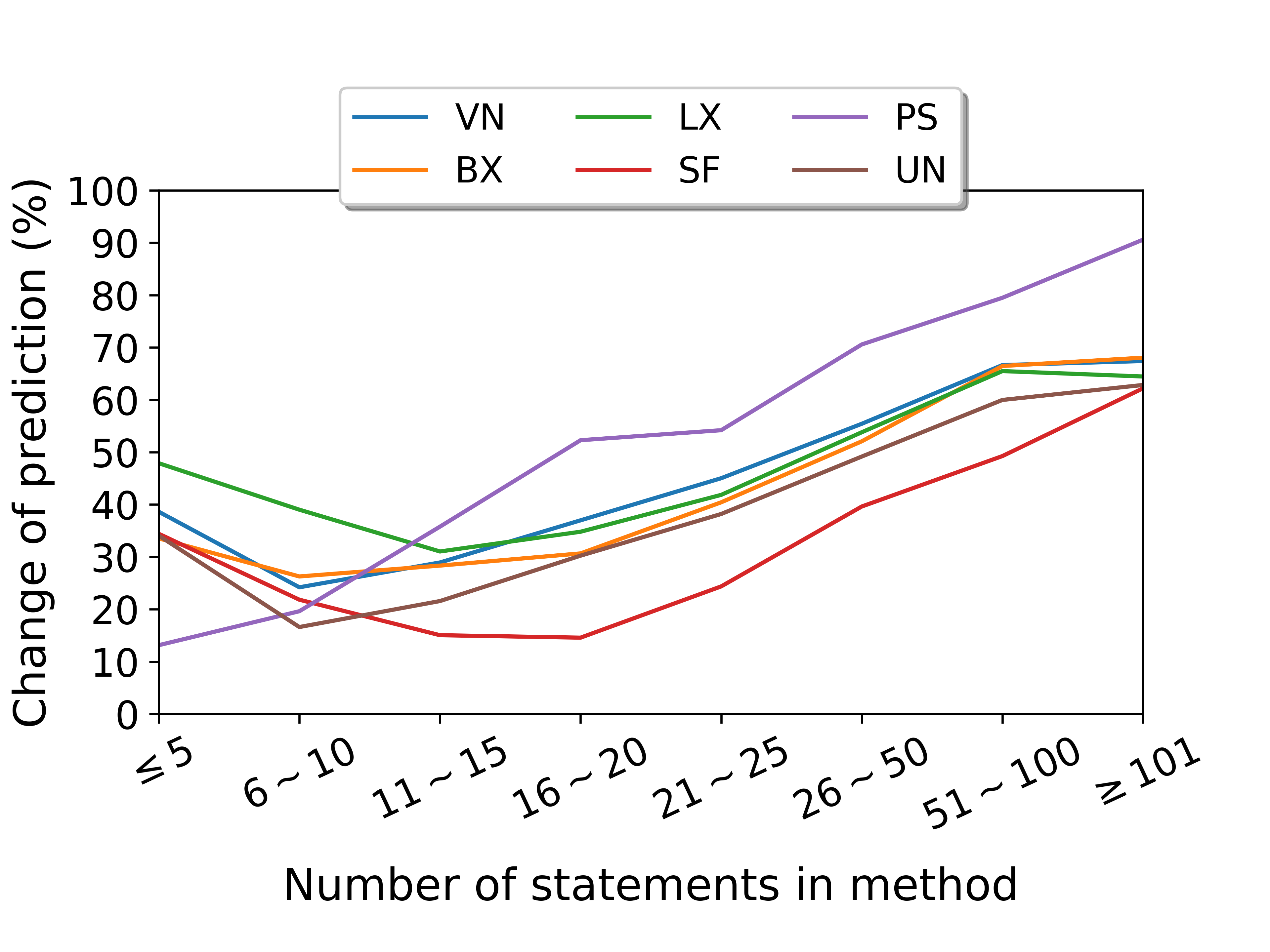}
\end{minipage}
\noindent \begin{minipage}{.33\textwidth}
\caption*{(d) \cts (\JS)}
\includegraphics[width=0.99\linewidth]{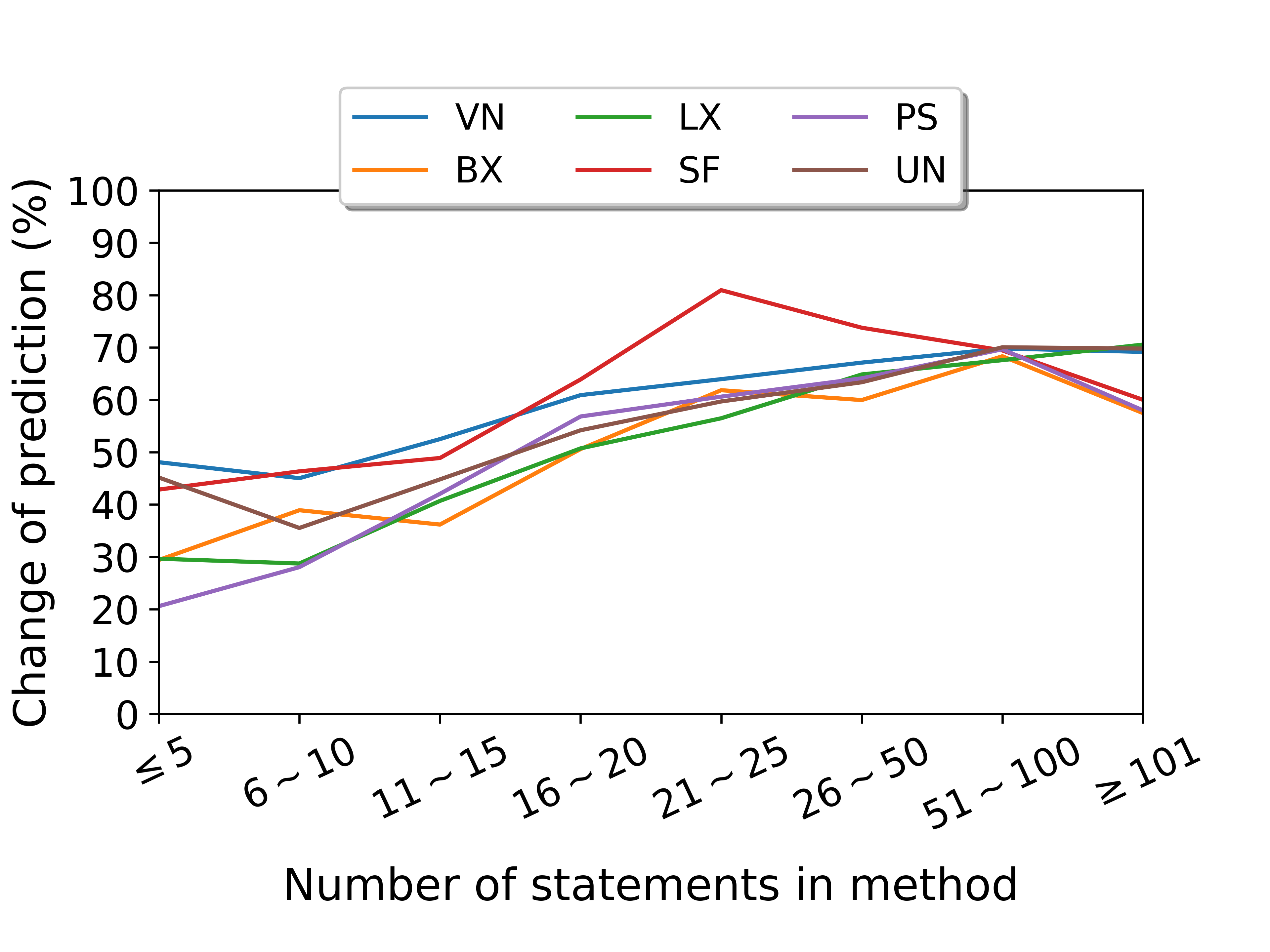}
\end{minipage}%
\begin{minipage}{.33\textwidth}
\caption*{(e) \cts (\JM)}
\includegraphics[width=0.99\linewidth]{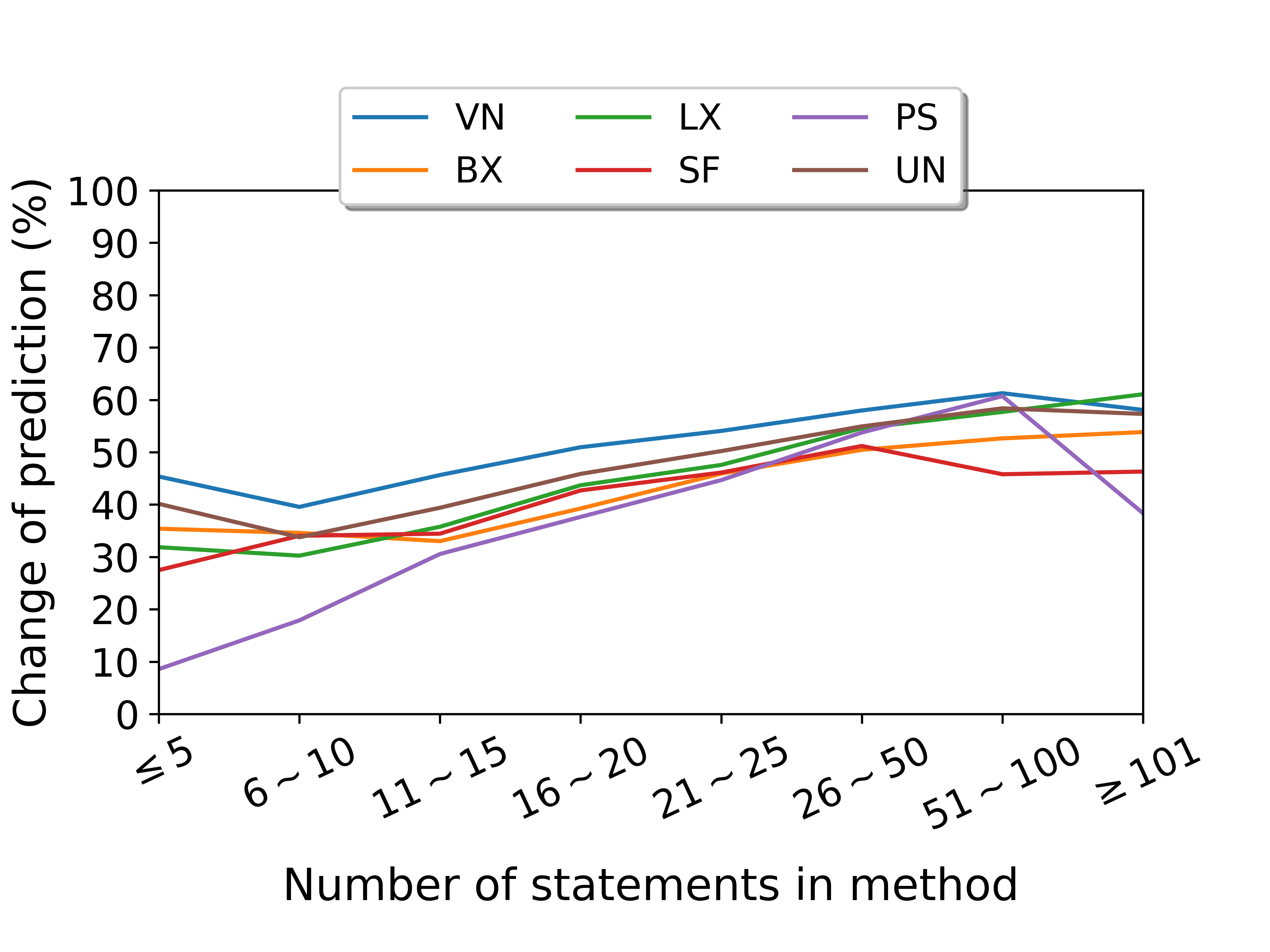}
\end{minipage}%
\begin{minipage}{.33\textwidth}
\caption*{(f) \cts (\JL)}
\includegraphics[width=0.99\linewidth]{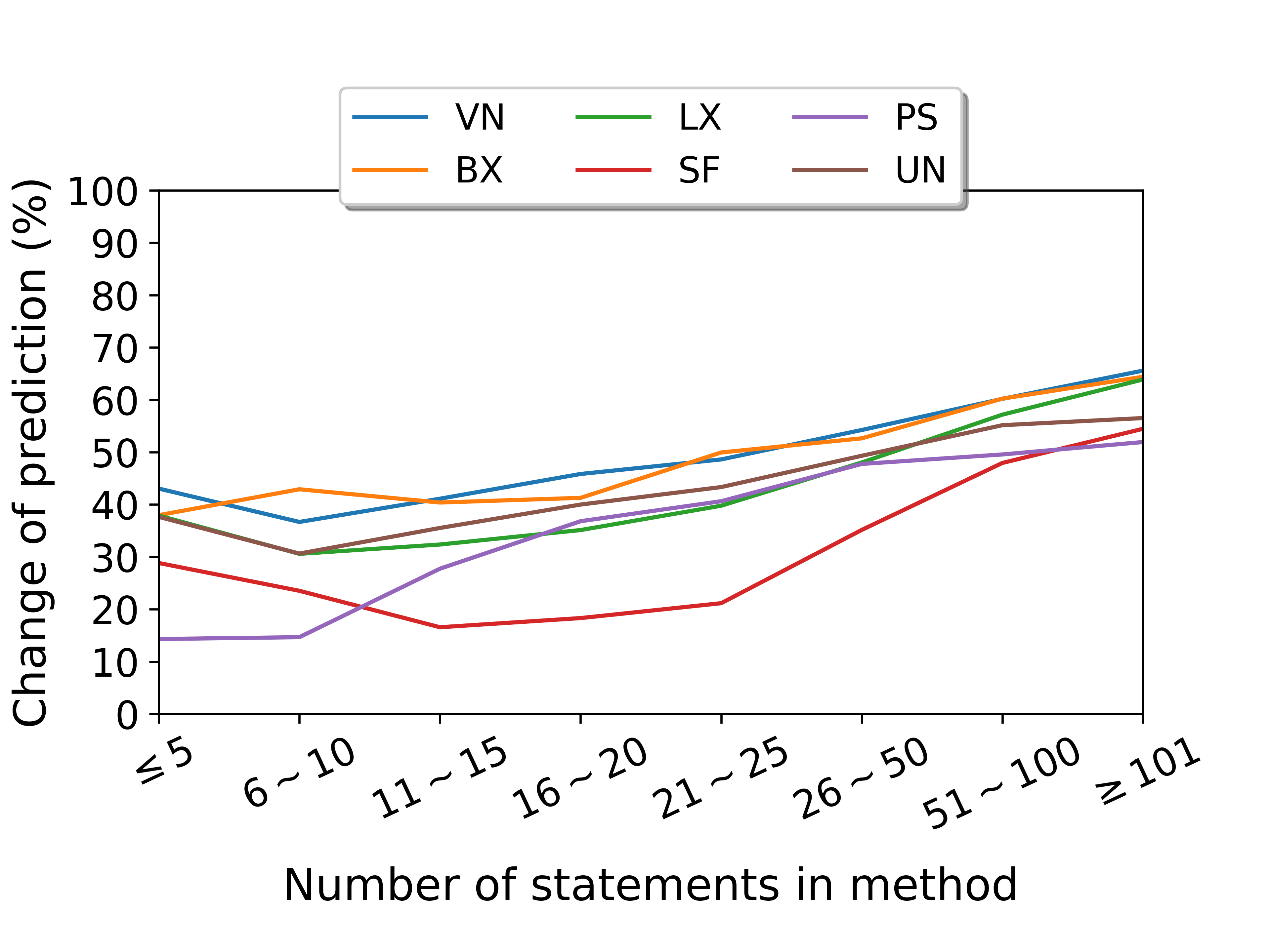}
\end{minipage}

\noindent \begin{minipage}{.33\textwidth}
\caption*{(g) \ggnn (\JS)}
\includegraphics[width=0.99\linewidth]{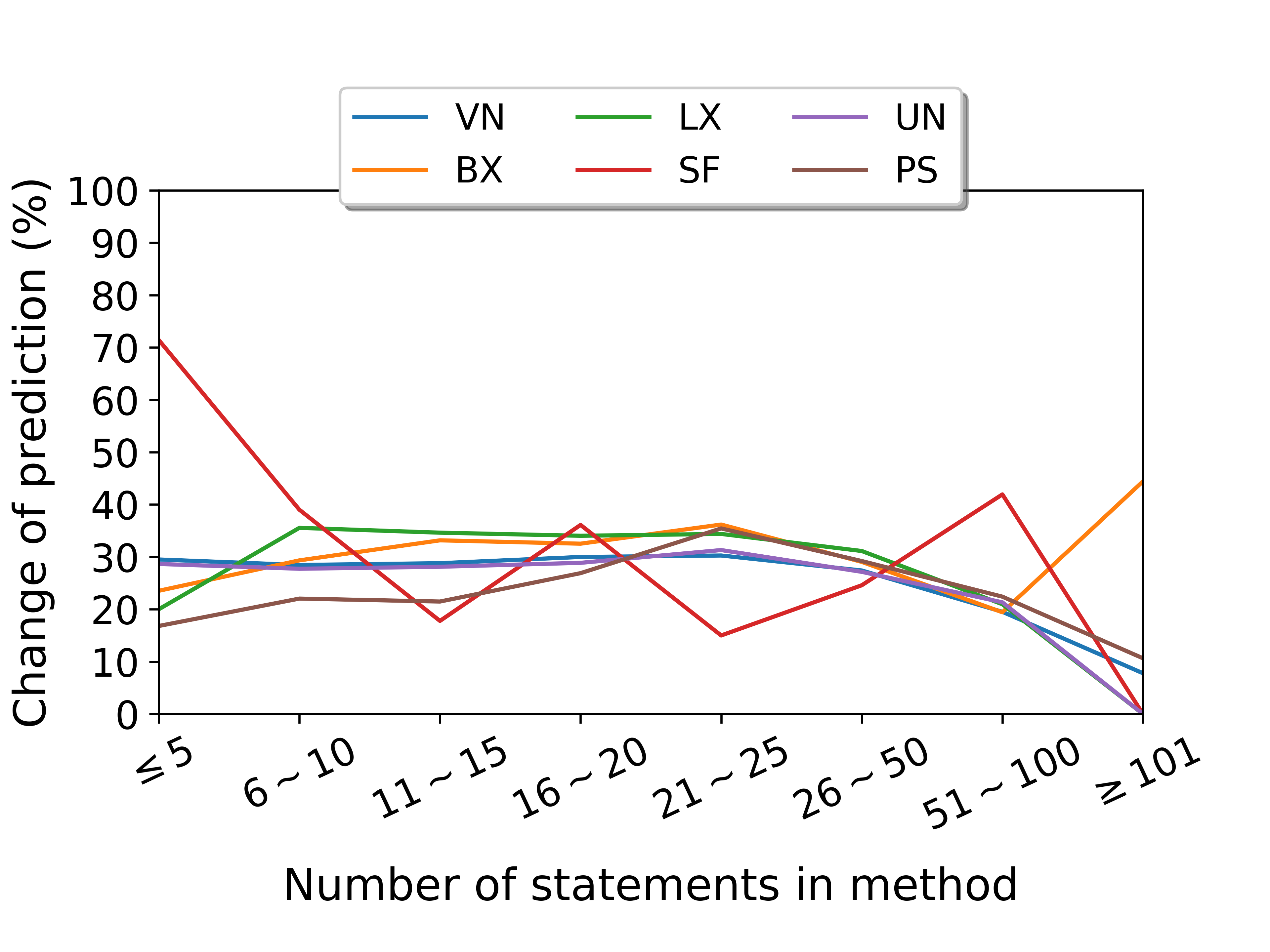}
\end{minipage}%
\begin{minipage}{.33\textwidth}
\caption*{(h) \ggnn (\JM)}
\includegraphics[width=0.99\linewidth]{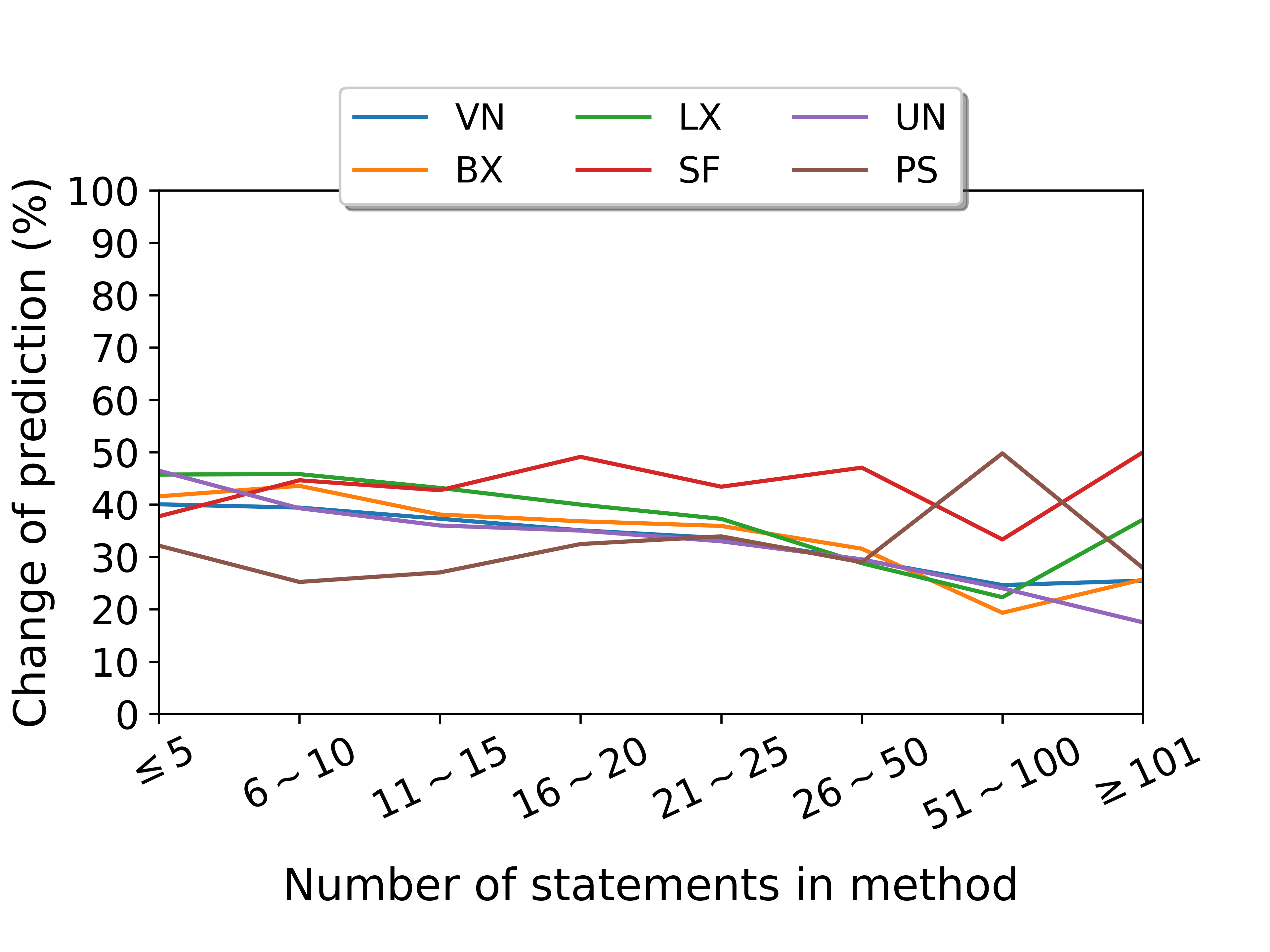}
\end{minipage}%
\begin{minipage}{.33\textwidth}
\caption*{(i) \ggnn (\JL)}
\includegraphics[width=0.99\linewidth]{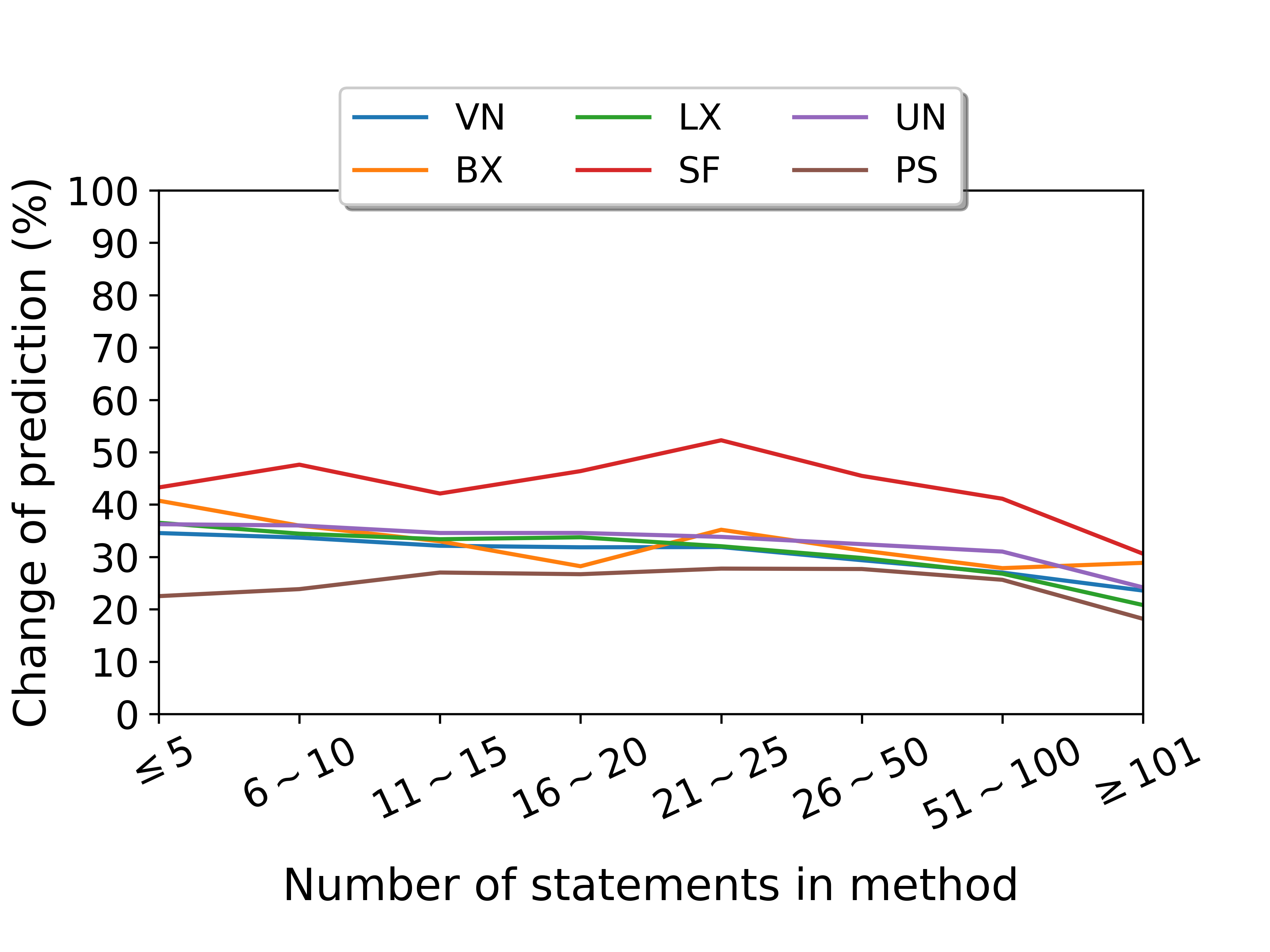}
\end{minipage}

\caption{Prediction Change Percentage (PCP) across the number of statements in methods.}
\label{fig:num_of_stmt}
\end{figure*}

An important metric of interest might be the generalizability in terms of the number of statements in the methods.
Figure~\ref{fig:num_of_stmt} depicts the relation between the length of methods and the prediction changes percentage (\ie, PCP) in the single-place transformed data.
In the figure, the ``Number of statements in method'' denotes the number of executable lines in the body of methods before the transformation.
%The trend in the figure suggests that there is a direct correlation between the size of programs and the percentage of prediction changes after transformation. 

As shown in Figure \ref{fig:num_of_stmt}(a-f), in most cases, the \ctv and \cts models exhibit notable increases in PCP for all the transformations and datasets as the number of lines in methods increases. However, looking at Figure~\ref{fig:num_of_stmt}(g-i), it seems that \ggnn is less sensitive to the number of lines in methods compared to \ctv and \cts with respect to the transformations.

%Therefore, scalability reveals that the number of statements in methods is a good indicator to emphasis the effectiveness of transformation, \eg the change is comparatively less sensitive on methods having few statements (\eg $\leq 5$), it is more severe on methods having many statements (\eg $\geq 501$).

\observation{The \ctv and \cts show notable increases in PCP as the length of methods grows, but PCP in \ggnn seems to be less sensitive to the length of methods.}

\subsection{RQ4: Trends in the Types of Changes}
%\begin{landscape}
\begin{table*}
  \begin{center}
    \caption{The detailed PCP across all models, datasets, and transformations.}
    \label{tab:cop_detailed}
    \def\arraystretch{1.25}
    \resizebox{\textwidth}{!}{%
    \begin{tabular}
    {|c|c|r|r|r|r|r|r|r|r|r|r|r|r|r|r|r|}
    \hline
    \multirow{2}{*}{\textbf{Dataset}} 
    & \multirow{2}{*}{\textbf{Transformation}} 
    & \multicolumn{3}{c|}{\textbf{CCP}} 
    & \multicolumn{3}{c|}{\textbf{CWP}}
    & \multicolumn{3}{c|}{\textbf{WWSP}}
    & \multicolumn{3}{c|}{\textbf{WCP}}
    & \multicolumn{3}{c|}{\textbf{WWDP}} \\ \cline{3-17}
    & 
    & \ctv & \cts & \ggnn
    & \ctv & \cts & \ggnn
    & \ctv & \cts & \ggnn
    & \ctv & \cts & \ggnn
    & \ctv & \cts & \ggnn
    \\ \hline \hline
    
    \multirow{6}{*}{\JS}
	    & \VN & 2.32 & 3.75 & { 15.76} & 1.18 & 1.65 & { 0.59} & 42.76 & 39.09 & { 56.08} & 0.57 & 1.21 & { 0.54} & 53.17 & 54.30 & { 27.03} \\ \cline{2-17}
	    & \BX & 3.88 & 4.54 & { 22.25} & 0.72 & 2.57 & { 0.43} & 42.26 & 41.15 & { 48.38} & 1.38 & 1.65 & { 0.58} & 51.76 & 50.09 & { 28.36} \\ \cline{2-17}
	    & \LX & 1.86 & 4.24 & { 16.23} & { 0.74} & { 0.74} & 0.89 & 38.76 & { 43.22} & 52.11 & 0.72 & 1.10 & { 0.62} & 57.92 & 50.70 & { 30.15} \\ \cline{2-17}
	    & \SF & 1.54 & 2.70 & { 26.61} & { 1.16} & 1.54 & 1.61 & 29.73 & 35.52 & { 41.94} & { 0.00} & 1.93 & 0.40 & 67.57 & 58.31 & { 29.44} \\ \cline{2-17}
	    & \PS & 4.64 & 3.96 & { 16.86} & 1.35 & 1.28 & { 0.80} & 22.57 & 38.72 & { 56.78} & 1.93 & 1.27 & { 0.27} & 69.51 & 54.77 & { 25.29} \\ \cline{2-17}
	    & \UN & 8.08 & 10.40 & { 20.97} & 1.73 & 3.20 & { 0.82} & { 51.96} & 44.00 & 50.69 & { 0.60} & 1.40 & 0.79 & 37.63 & 41.00 & { 26.73} \\ \hline
    \hline 
    \multirow{6}{*}{\JM}
	    & \VN & 7.56 & 9.41 & { 20.39} & 2.81 & 3.90 & { 1.18} & { 45.89} & 41.85 & 43.65 & { 0.80} & 1.54 & 1.22 & 42.94 & 43.30 & { 33.56} \\ \cline{2-17}
	    & \BX & 12.76 & 13.90 & { 27.83} & 1.79 & 1.91 & { 1.00} & 36.89 & { 41.39} & 38.42 & 1.45 & 1.97 & { 1.03} & 47.11 & 40.83 & { 31.72} \\ \cline{2-17}
	    & \LX & 6.61 & 7.62 & { 20.93} & 2.21 & 1.71 & { 1.09} & 30.62 & { 47.09} & 42.40 & { 1.22} & 1.33 & 1.23 & 59.34 & 42.25 & { 34.35} \\ \cline{2-17}
	    & \SF & 11.15 & 17.90 & { 31.28} & 5.99 & 2.94 & { 2.93} & 28.94 & { 40.51} & 24.99 & 2.45 & 2.27 & { 1.85} & 51.47 & { 36.38} & 38.95 \\ \cline{2-17}
	    & \PS & 11.53 & 14.55 & { 25.47} & 4.31 & 1.57 & { 1.09} & 23.03 & { 42.81} & 40.44 & 2.05 & 1.57 & { 1.29} & 59.08 & 39.50 & { 31.71} \\ \cline{2-17}
	    & \UN & 16.90 & 21.58 & { 25.07} & 3.24 & 5.22 & { 1.83} & { 47.30} & 38.18 & 32.13 & { 0.79} & 1.41 & 1.67 & { 31.77} & 33.61 & 39.30 \\ \hline
    \hline 
    \multirow{6}{*}{\JL}
	    & \VN & { 15.40} & 14.54 & 14.57 & 3.25 & 4.69 & { 0.77} & 42.55 & 38.42 & { 53.51} & 1.40 & 2.32 & { 1.13} & 37.40 & 40.03 & { 30.02} \\ \cline{2-17}
	    & \BX & 11.33 & 10.14 & { 13.97} & 2.09 & 2.92 & { 0.79} & 40.87 & 38.42 & { 54.06} & 2.00 & 2.92 & { 1.13} & 43.71 & 45.60 & { 30.05} \\ \cline{2-17}
	    & \LX & { 19.81} & 18.87 & 13.29 & 3.00 & 2.49 & { 0.28} & 33.67 & 38.62 & { 54.96} & 2.13 & 2.30 & { 1.13} & 41.39 & 37.72 & { 30.34} \\ \cline{2-17}
	    & \SF & 48.30 & { 52.23} & 9.30 & 5.63 & 5.28 & { 0.27} & 21.37 & 18.68 & { 45.20} & 4.09 & 2.80 & { 0.68} & 20.61 & 21.01 & { 44.55} \\ \cline{2-17}
	    & \PS & 11.13 & 13.35 & { 21.89} & 6.66 & 2.17 & { 1.27} & 24.49 & 44.72 & { 51.79} & 2.89 & 2.82 & { 0.96} & 54.83 & 36.94 & { 24.09} \\ \cline{2-17}
	    & \UN & 24.31 & { 26.57} & 22.28 & 3.75 & 5.95 & { 1.19} & { 44.47} & 35.99 & 42.05 & { 0.85} & 1.81 & 1.24 & { 26.62} & 29.68 & 33.24 \\ \hline
  
    \end{tabular}%
    }
  \end{center}
\end{table*}
%\end{landscape}

Table~\ref{tab:cop_detailed} shows the full breakdown of the proportion of different types of changes after the transformation of methods.
In this experiment, we use the same single-place transformed data that have been used for the PCP in Table \ref{table:cop_summary}.
In \ctv and \cts, the value of CCP increases with increase in the size of datasets. It may suggest that with a larger dataset the \npa can generalize the correct predictions better.

In addition, we calculate $\frac{CWP}{CCP+CWP}$ to approximate the ratio of cases that the \npa's prediction switches from correct to wrong after transformations with respect to all the cases whose initial predictions are correct.
The ratio helps us to simplify the comparison of (in)generalizability across different models.
On average, 23\% and 20\% of cases, the \npa switches from a correct prediction to a wrong one in \ctv and \cts, respectively. In \ggnn, on the other hand, this switch happens in less than 5\% of transformations.

Similarly, $\frac{WCP}{WWSP+WWDP+WCP}$ approximates the ratio of cases switching from a wrong prediction to a correct prediction after transformations with respect to all the cases whose initial prediction are wrong. 
In \ctv and \cts, a transformation switches from a wrong prediction to correct prediction in less than 3\% of cases, however, this switch happens in around 1\% of transformations for \ggnn.
Higher $\frac{CWP}{CCP+CWP}$ than $\frac{WCP}{WWSP+WWDP+WCP}$ implies that transformations are likely to reduce the overall performance of the \npas.

\observation{Transformations are likely to decrease the overall performance of \npas, and they are more likely to change the correct prediction in \ctv and \cts than \ggnn, while the generalizability of \ctv and \cts can be compensated by larger datasets more than \ggnn.}

\subsection{RQ5: Impact of the Transformations on Precision, Recall, and \texorpdfstring{$F_1$-Score}{}}
\begin{table*}[t]
    \caption{The precision, recall and $F_1$-score for subtokens across all models, datasets, and transformations.}
    \label{table:prf_summary}
    \def\arraystretch{1.25}
    \resizebox{0.9\textwidth}{!}{%
    \begin{tabular}{|c|c|r|r|r|r|r|r|r|r|r|r|}
        \hline
        \multirow{2}{*}{\textbf{Dataset}}
        & \multirow{2}{*}{\textbf{Transformation}}
        & \multirow{2}{*}{\textbf{\pbox{10cm}{\# Transformed \\ methods}}}
        & \multicolumn{3}{c|}{\textbf{Precision}} 
        & \multicolumn{3}{c|}{\textbf{Recall}} 
        & \multicolumn{3}{c|}{\textbf{$F_1$-Score}} \\ \cline{4-12}
        & & & \textbf{\ctv} & \textbf{\cts} & \textbf{\ggnn}
        & \textbf{\ctv} & \textbf{\cts} & \textbf{\ggnn}
        & \textbf{\ctv} & \textbf{\cts} & \textbf{\ggnn} \\
        \hline \hline
        
        \multirow{7}{*}{\JS}
        & \VN & 123123 & 9.79 & 38.01 & {\bf 40.64} & 5.05 & {\bf 28.99} & 23.78 & 6.66 & {\bf 32.89} & 30.00 \\  \cline{2-12}
        & \BX & 1519 & 8.97 & 33.58 & {\bf 41.19} & 5.36 & {\bf 26.94} & 25.93 & 6.71 & 29.90 & {\bf 31.83} \\  \cline{2-12}
        & \LX & 5160 & 9.08 & 34.52 & {\bf 39.50} & 5.22 & {\bf 26.08} & 23.40 & 6.63 & {\bf 29.71} & 29.39 \\  \cline{2-12}
        & \SF & 259 & 7.01 & 30.78 & {\bf 38.32} & 4.99 & 26.41 & {\bf 30.16} & 5.83 & 28.43 & {\bf 33.75} \\  \cline{2-12}
        & \PS & 9169 & 11.21 & 33.11 & {\bf 45.73} & 5.64 & {\bf 25.52} & 22.95 & 7.50 & 28.82 & {\bf 30.56} \\  \cline{2-12}
        & \UN & 44426 & 21.26 & {\bf 50.99} & 44.37 & 13.63 & {\bf 41.12} & 31.36 & 16.61 & {\bf 45.53} & 36.75 \\  \cline{2-12}
        & \multicolumn{2}{r|}{Weighted Average =} & 12.60 & 40.76 & {\bf 41.77} & 7.16 & {\bf 31.65} & 25.59 & 9.11 & {\bf 35.62} & 31.66 \\
        \hline \hline 
        
        \multirow{7}{*}{\JM}
        & \VN & 771208 & 20.90 & {\bf 43.57} & 39.71 & 10.89 & {\bf 28.98} & 22.24 & 14.32 & {\bf 34.81} & 28.51 \\  \cline{2-12}
        & \BX & 8840 & 22.29 & 40.72 & {\bf 42.26} & 13.95 & {\bf 29.02} & 24.72 & 17.16 & {\bf 33.89} & 31.19 \\  \cline{2-12}
        & \LX & 23533 & 18.29 & 39.25 & {\bf 42.00} & 10.06 & {\bf 26.55} & 23.10 & 12.98 & {\bf 31.67} & 29.81 \\  \cline{2-12}
        & \SF & 3839 & 30.24 & {\bf 51.49} & 47.67 & 20.89 & {\bf 39.56} & 34.65 & 24.71 & {\bf 44.74} & 40.13 \\  \cline{2-12}
        & \PS & 44711 & 24.29 & 38.75 & {\bf 39.79} & 12.99 & {\bf 28.26} & 23.35 & 16.93 & {\bf 32.68} & 29.43 \\  \cline{2-12}
        & \UN & 351621 & 32.87 & {\bf 55.79} & 44.11 & 21.56 & {\bf 43.15} & 30.03 & 26.04 & {\bf 48.66} & 35.73 \\  \cline{2-12}
        & \multicolumn{2}{r|}{Weighted Average =} & 24.51 & {\bf 46.88} & 41.09 & 14.12 & {\bf 33.08} & 24.63 & 17.87 & {\bf 38.74} & 30.74 \\
        \hline \hline 
        
        \multirow{7}{*}{\JL}
        & \VN & 916565 & 35.17 & {\bf 48.79} & 32.80 & 20.83 & {\bf 38.14} & 20.46 & 26.16 & {\bf 42.81} & 25.20 \\  \cline{2-12}
        & \BX & 12107 & 27.30 & {\bf 42.90} & 23.82 & 15.83 & {\bf 33.19} & 16.54 & 20.04 & {\bf 37.43} & 19.52 \\  \cline{2-12}
        & \LX & 49665 & 37.60 & {\bf 46.79} & 25.28 & 23.91 & {\bf 37.91} & 18.75 & 29.23 & {\bf 41.88} & 21.53 \\  \cline{2-12}
        & \SF & 11165 & 69.34 & {\bf 72.75} & 22.68 & 57.06 & {\bf 66.18} & 21.81 & 62.60 & {\bf 69.31} & 22.24 \\  \cline{2-12}
        & \PS & 74973 & 25.56 & {\bf 43.88} & 32.56 & 14.62 & {\bf 33.28} & 21.37 & 18.60 & {\bf 37.85} & 25.80 \\  \cline{2-12}
        & \UN & 370927 & 44.96 & {\bf 61.57} & 45.43 & 30.40 & {\bf 52.44} & 29.93 & 36.27 & {\bf 56.64} & 36.09 \\  \cline{2-12}
        & \multicolumn{2}{r|}{Weighted Average =} & 37.48 & {\bf 51.90} & 35.64 & 23.32 & {\bf 41.75} & 22.87 & 28.72 & {\bf 46.25} & 27.85 \\
        \hline

        \end{tabular}%
        }
\end{table*}

The performance of \mbox{\npas} in the literature are often measured in classic metrics, such as precision, recall, and $F_1$-score.
In particular for the method name prediction task trained for \ctv, \mbox{\cts} and \ggnn, subtoken-level comparison is used to calculate the metrics; i.e., the method names in both predicted results and ground-truth names are split into individual tokens for the measurements (cf.~the definitions in Section~\mbox{\ref{sec:metrics}}).

We also study the impact of the program transformations on the performance of \mbox{\npas} in terms of these classic metrics.
Table \mbox{\ref{table:prf_summary}} shows the changed precision, recall, and $F_1$-scores for the programs transformed by different transformations.
In this experiment, we use the same single-place transformed data that have been used for the PCP in Table \mbox{\ref{table:cop_summary}}.

In comparison with Table \mbox{\ref{table:all_models}}, we can see the average precision, recall, and $F_1$-score in Table \mbox{\ref{table:prf_summary}} have obvious decreases for all the three \mbox{\npas} across the three Java datasets, that may indicate the (negative) impact of the transformations on the \mbox{\npas}.

We also find no obvious correspondence between the PCP shown in Table \mbox{\ref{table:cop_summary}} and the changes in precision, recall, and $F_1$-score; high PCP does not necessarily lead to high changes in precision, recall and $F_1$-score and vice versa.

\observation{\mbox{\Npas} seem susceptible to semantic-preserving transformations with respect to the classic metrics of precision, recall, and $F_1$-score as well. While our new metric of Prediction Change Percentage (PCP) shows the impact of the transformations from a different and more fine-grained perspective, the changes in the classic metrics are not correlated with PCP.}

\section{Discussion}
\label{sec:discuss}

In this paper, we study the current state of generalizability in \npas built on \ctv, \cts, and \ggnn. Although limited, it provides interesting insights. In this section, we discuss why neural networks have become a popular, or perhaps the de-facto, tool for processing programs, and what are the implications of using neural networks in processing source code.

\begin{comment}
    We apply transformations on a base program to generate new programs and use a trained model (\eg \ctv, \cts, \ggnn) to make a prediction on base program and each generated program.
    During training, the models learn the embeddings of terminals, paths, and tags with other network parameters.
    During prediction, models extract the bag of AST paths from given code snippet and select corresponding embeddings from the lookup table.
    If the transformations delete, insert or modify the value or position of nodes in AST, different embeddings can be returned from the lookup table.
    As a result, a different context vector will be created and different attention will be given. 
    Therefore, it might end up creating different code vector and predicting different tags as well.

\end{comment}

%\subsection{\ke{This section is confusing to me. Why do we need it?} NN Blessings: hypothesis class and feature learning}
%The applicability of machine learning techniques depends on the hypothesis class they represent. That is, the decision boundary that can be represented by the technique. For instance, linear regression can effectively be used for linearly separable problems, but they are not as effective in problems with higher complexity.

Neural networks constitute a powerful class of machine learning models with a large hypothesis class. For instance, a multi-layer feed-forward network is called a universal approximator, meaning that it can essentially represent any function~\cite{hornik1989multilayer}. Unlike traditional learning techniques that require extensive feature engineering and tuning, deep neural networks facilitate representation learning.
That is, they are capable of performing feature extraction out of raw data on their own~\cite{lecun2015deep}.
Given a sufficiently large dataset, neural networks with adequate capabilities can substantially reduce the burden of feature engineering. 
Availability of a large number of code repositories makes data-driven program analysis a good application of neural networks. However, it is still unknown if neural networks are the best way to process programs~\cite{Devanbu:FSE:2017} vs.~\cite{Sutton:2019:maybe}.

Although the large hypothesis class of neural networks and feature learning make them very appealing to use, the complex models built by neural networks are still too difficult to understand and interpret.
Therefore, as we apply neural networks in program analysis, we should develop specialized tools and techniques to enhance their interpretability, generalizability and robustness.
%\Fix{LX: the above is about interpretability; but we're about generalizability...I added a few sentences in the subsection below to contrast them.}

\subsection{Generalizability vs.~Interpretability vs.~Robustness and Others}
Interpretability studied in the literature may help to build more understandable neural networks, revealing the limits and strengths of the networks, and thus to some extent, it helps to evaluate and understand the generalizability of the networks. However, our study of generalizability with respect to program transformation provides a different perspective complement to interpretability; the approach may have the potential in the future to help identify interpretable code elements by measuring the impact of certain types of code transformations.

%\Fix{Rabin: same as the footnote of page 2?}
As mentioned in Section~\ref{sec:example}, there is a substantial line of work on evaluating the robustness of neural networks especially in the domain of vision and pattern recognition~\cite{szegedy2013intriguing}. The key insight in such domains is that small, imperceptible changes in input should not impact the result of output. While this observation can be true for domains such as vision, it might not be directly applicable to the discrete domain of \npas, since some minor changes to a program can drastically change the semantic and behavior of the program.
Quantifying the imperceptibility and many other aspects of source code is our future research goal.

\subsection{Are we there yet?}
Are \npas ready for widespread use in program analysis? 
The \npas in our experiments are brittle to even very small changes in the methods. 
The semantic-preserving transformations can change the outputs of the \npas in 26\% to 73\% of cases. 
Although our findings are limited to only one task, they suggest caution. 
The literature lacks techniques for rigorous evaluation of \npas.
The recent line of work by Nghi et al. 
\cite{Nghi2019AutoFocus} in interpretability of \npas, Rabin et al. \cite{rabin2019tnpa, rabin2020demystifying,rabin2020evaluation} in testing them, and Yefet et al. \cite{yefet2019adversarial} are much needed steps in a right direction.

\subsection{Code Representation}
The performance of models used in \npas, such as ones used in this study, is relatively low compared to the performance of neural models in domains such as natural language understanding~\cite{sarikaya2014application}, text classification~\cite{lai2015recurrent}.
To improve their performance, we would need novel code representations that better capture interesting characteristics of programs. 

\section{Related Work}
\label{sec:related}

\Part{Robustness of Neural Networks.}
There is a substantial line of work on the robustness of artificial intelligence (AI) systems in general and deep neural networks in particular.~\citet{szegedy2013intriguing} is the first to discover that deep neural networks are vulnerable to small perturbations that are imperceptible to human eyes. They developed the L-BFGS method for the systematic generation of such adversarial examples.~\citet{goodfellow2014explaining} proposes a more efficient method, called the Fast Gradient Sign Method that exploits the linearity of deep neural networks. Many following up works \mbox{\cite{kurakin2016adversarial, moosavi2016deepfool, carlini2017towards, dong2018boosting, yuan2019adversarial, zhang2020gnnguard, yefet2019adversarial}} further demonstrated the severity of the robustness issues with a variety of attacking methods and defenses. While aforementioned approaches only apply to models for image classification, new attacks have been proposed that target models in other domains, such as natural language processing~\cite{li2016understanding,jia2017adversarial,zhao2018generating} and graphs~\cite{Dai_Li_Tian_Huang_Wang_Zhu_Song_2018,zugner2018adversarial}.

The automated verification research community has proposed techniques to offer guarantees for the robustness of neural networks by adapting bounded model checking~\cite{scheibler2015towards}, abstract interpretation~\cite{gehr2018ai2}, and Satisfiability Modulo Theory~\cite{HX}.
\citet{zimmermann2019SE4ML} study the challenges in developing AI solutions and
\citet{harman2020TestingSurvey} survey testing of machine-learning systems.

% \Part{Adversarial Examples}
% Adversarial examples for a neural model are those that can mislead the model to mispredict an input.

\Part{Models of Code.}
Early works directly adopted NLP models to discover textual patterns existed in the source code~\cite{gupta2017deepfix,pu2016sk_p}. Those methods, unfortunately, do not account for the structural information programs exhibit. Following approaches address this issue by generalizing from the abstract syntax trees~\cite{maddison2014structured,mou2016convolutional,alon2018code2vec,alon2018code2seq}. 
As Graph Neural Networks (GNN) \cite{scarselli2009ggn} have been gaining increasing popularity due to its remarkable representation capacity, many works have leveraged GNN to tackle challenging tasks like program repair and bug finding, and obtained quite promising results~\cite{allamanis2017learning,wang2019learning,dinella2019hoppity, li2015gated, BrockschmidtAGP19, dinella2019hoppity, tarlow2020learning}.
Besides, the attention mechanism \mbox{\cite{bahdanau2015attention}} has been applied into GNNs to improve the performance further \mbox{\cite{beck2018g2s,xu2018graph2seq,ZhangX20}}. It is very interesting to see how the attention can help to explain the output of the neural models \mbox{\cite{VelickovicCCRLB18,xu2018modeling,Nghi2019AutoFocus}}.
In parallel, Wang ~\etal~ developed a number of models~\cite{wang2017dynamic,wang2018dynamic,wang2020semantic} that feed off the run time information for enhancing the precision of semantic representation for model inputs.

\section{Threats to Validity}
\label{sec:threats}
There are various threats to the validity of our approach.

\Part{Limited Data and Evaluation Scope.}
We only evaluated the generalizability of \npas built on \ctv, \cts, and \ggnn, for one task in Java programs. Therefore, our results may not generalize to other \npas or other tasks or other programming languages. We leave the evaluation of the general applicability of our approach as future work.

\Part{Transformations.}
The proposed transformations in this paper impact program ASTs in varying degrees. Some of the transformations, e.g. variable renaming, are common refactoring techniques. However, these transformations may not represent many possible transformations in other domains. We will instantiate and extend our approach with other transformations from other domains. 

\Part{Internal Validity.}
Some bugs may exist in the toolchain and \npas implemented in this paper.
To reduce the probability of bugs, two authors reviewed the code and manually inspected a sample of transformed programs to ensure the reliability of transformations.

\section{Conclusion \& Future Work}
\label{sec:conclude}

In this paper, we perform a large-scale, systematic evaluation of the generalizability of state-of-the-art \npas built on \ctv, \cts, and \ggnn.
In particular, we apply six semantic-preserving program transformations to produce new programs on which we expect the \npas to keep their original predictions. 
We find that such program transformations frequently sway the predictions of these \npas, indicating serious generalization issues that could negatively impact the wider applications of deep neural networks in program analysis tasks.
Although \npas that encode more program dependency information and are trained with larger datasets may exhibit more generalizable behavior, their generalizability is still limited. 
We believe this work provides a systematic approach and metrics for evaluating \npas, and can motivate future research on training not only accurate but also generalizable deep models of code.
Future work that includes more semantic-preserving and even some semi-semantic-preserving transformations in our approach and adapts more fine-grained prediction change metrics may further extend the applicability of our approach to various \npas designed for different tasks. 
We also plan to explore using transformed programs to improve the generalizability of the \npas.

\section*{Acknowledgments}

This research is supported by the Singapore Ministry of Education (MOE) Academic Research Fund (AcRF) Tier 1 Grant No. 19-C220-SMU-002 and the Research Lab for Intelligent Software Engineering (RISE) Operational Fund from the School of Computing and Information Systems (SCIS) at Singapore Management University (SMU). We also thank the anonymous reviewers for their insightful comments and suggestions, and thank the authors of previous related works for sharing data and models.

\balance
\bibliographystyle{ACM-Reference-Format}
\bibliography{references}
\end{document}